\documentclass[twocolumn,secnumarabic,amssymb, nobibnotes, prx,superscriptaddress]{revtex4-1}

\usepackage{pgfplots}
\pgfplotsset{compat=1.18}
\usepackage{graphicx}
\usepackage{dcolumn}
\usepackage{amsmath} 
\usepackage{xcolor}
\usepackage{xparse}
\usepackage{mathtools}

\usepackage[T1]{fontenc}
\usepackage{libertine}
\usepackage{relsize}
\usepackage{bm}
\usepackage[caption=false]{subfig}

\usepackage{amssymb}
\usepackage{natbib}
\usepackage{mathrsfs}
\usepackage[T1]{fontenc}

\begin{document}

\title{Formation of Topologically Associated Chromatin Domains\\ using Quantum Annealing}

\author{Tobias Kempe}
\email{tkempe96@gmail.com}
\affiliation{Department of Physics, RWTH Aachen University, Aachen, Germany}
 
\author{S.M. Ali Tabei}%
\email{tabei@uni.edu}
\thanks{Corresponding author}
\affiliation{Department of Physics, University of Northern Iowa, Cedar Falls, Iowa 50614, United States}%

\author{Mohammad H. Ansari}
\email{m.ansari@fz-juelich.de}
\thanks{Corresponding author}
\affiliation{Peter Gr\"unberg Institute (PGI-12), Forschungszentrum Jülich, J\"ulich 52428, Germany}

\date{\today}

\begin{abstract}
Topologically Associating Chromatin Domains are spatially distinct chromatin regions that regulate transcription by segregating active and inactive genomic elements. Empirical studies show that their formation correlates with local patterns of epigenetic markers, yet the precise mechanisms linking 1D epigenetic landscapes to 3D chromatin folding remain unclear. Recent models represent chromatin as a spin system, where nucleosomes are treated as discrete-state variables coupled by interaction strengths derived from genomic and epigenetic data. Classical samplers struggle with these models due to high frustration and dense couplings. Here, we present a quantum annealing (QA) approach to efficiently sample chromatin states, embedding an epigenetic Ising model into the topology of D-Wave quantum processors. Rather than reconstructing exact TAD size distributions or insulation scores, our method reproduces statistical features, such as mean marker incidences and intra-/inter-nucleosome correlations, while generating configurations that exhibit TAD-like structural motifs. These results demonstrate QA as an alternative to explore the chromatin architecture and provide a foundation in epigenetic modeling.

\end{abstract}

\maketitle


\label{sec:introduction}
\section{Introduction}

The spatial organization of the genome plays a fundamental role in regulating gene expression and maintaining cellular identity. Advances in chromosome conformation capture techniques, such as Hi-C \cite{beltonHi-C, mccord2020chromosome}, have enabled genome-wide mapping of physical contacts between distant DNA loci, revealing that chromatin is folded into hierarchical structures including loops, compartments, and topologically associating domains (TADs) \cite{Lieberman2009, Rao20143d}. These structures reflect a complex interplay between DNA sequence, architectural proteins, and epigenetic modifications.
Epigenetics refers to heritable changes in gene expression that occur without altering the underlying DNA sequence. Epigenetic marks, including histone modifications and DNA methylation, act as chemical tags on chromatin. They help regulate gene expression and higher-order folding. Many of these marks are stable across cell divisions and can even be transmitted across generations \cite{DNAMethylationMemory,TransgenerationalEpigeneticInheritance}, yet they remain dynamic and responsive to environmental and developmental cues. Certain marks play key roles in a number of medical conditions, including cancer, autoimmune diseases, mental disorders, and diabetes \cite{egger2004epigenetics, jones2002Cancer, Jiang2004, quintero2012autoimmune, aslani2016autoimmune, nestler2016mental, ling2019diabetes}. 

\begin{figure*}[t]
\centering
\includegraphics[width=2 \columnwidth]{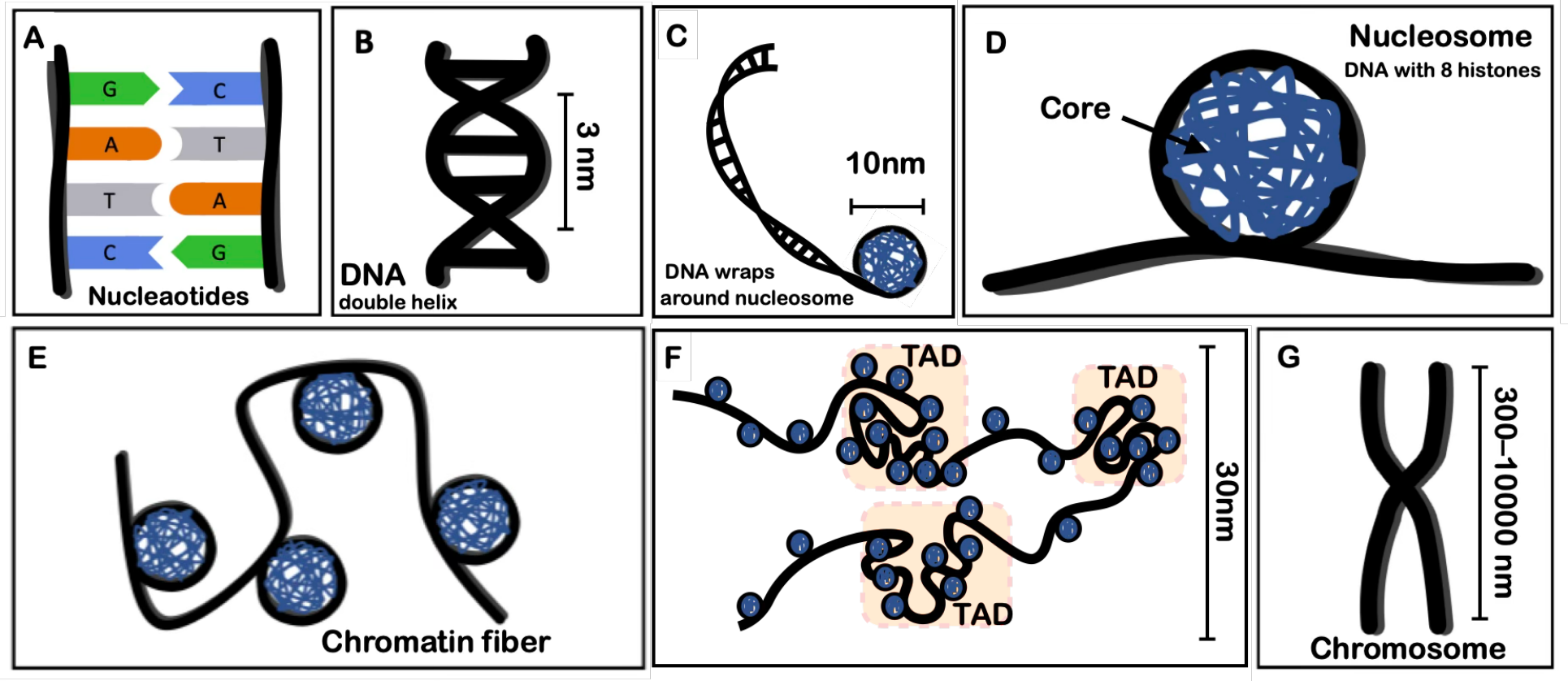}
\caption{Chromosome structure. (A) Nucleaotide base pairs. (B) DNA double helix. (C) Schematic illustration of DNA wrapping around a nucleosome (not to scale), with the nucleosome core shown in the schematic illustration of (D) being a (blue) DNA strand wrapped around histone proteins forming a nucleosome. (E) Euchromatin as beads on a string configuration of nucleosomes with spacer DNA strands in between. This is the form of chromatin that is modeled later in this paper. (F) Schematic formation of TADs in heterochromatin (compact euchromatin) as a result of interaction between nucleosomes. (G) Chromosomes, of which 23 pairs exist in the human genome. }
\label{fig:bio:chromatin_structure}
\end{figure*}

Furthermore, epigenetic marks can be altered by environmental factors and behavior \cite{roth2013behavior} such as diet \cite{mckay2011diet} and psychological well-being \cite{pluess2015wellbeing}, and can even be influenced by social interactions (e.g., maternal-infant bonding) \cite{MaternalInfantInteractionEpimarks}.

In addition to the modification of gene expression and interaction with phenotypical features, epigenetic marks can also influence the three-dimensional folding of chromatin.  Chromatin is an intermediate level of genome organization in the structural hierarchy between DNA base pairs and the full structure of chromosomes. It consists of DNA tightly packaged with histone and non-histone proteins, which mediate its three-dimensional folding and spatial organization within the nucleus of mostly eukaryotic cells \cite{EpigeneticsIntroduction}. Figure \ref{fig:bio:chromatin_structure} visualizes the structural hierarchy, i.e. ,the elements emerging towards a chromosome. 
Eukaryotic chromatin is spatially organized into topologically associating domains (TADs), which are kilobase- to megabase-sized regions where regulatory elements, histone modifications, and DNA methylation patterns co-align to promote coordinated gene regulation \cite{DNAMethylationMemory, ChromatinDomains}. By segregating transcriptionally active loci from repressed ones, these often conserved structural units fine-tune the probability and timing of gene expression and underlie higher-order genome architecture.

Deciphering TAD formation is critical for unraveling gene-regulatory networks, disease etiology, and prospective therapeutic strategies \cite{DomainFormationPathogenes}. However, the precise mechanisms linking 1D epigenetic landscapes to 3D chromatin folding remain unclear. Recent advances have explored this link using polymer-based and statistical models, including approaches that infer a combinatorial code connecting chromatin states to 3D architecture \cite{Esposito2022}. These models leverage polymer physics and machine learning to predict genome-wide contact maps from epigenetic features, complementing earlier frameworks based on phase separation and loop extrusion \cite{ComplexityChromatinFolding,Chiariello2020,Fudenberg2016}. Our work builds on these concepts by introducing a quantum annealing approach to efficiently sample chromatin configurations, offering a novel computational perspective on the physical principles underlying TAD formation.

To quantitatively model this process, chromatin is often abstracted as an Ising-like spin system, where nucleosomes are treated as discrete-state variables coupled by interaction strengths derived from genomic and epigenomic data \cite{CDFLearning, erdel2013establishing, katava2022chromatin}. Each nucleosome is assigned a discrete epigenetic ``spin,'' while coupling constants encode intra- and inter-nucleosome interactions. Domain configurations are then obtained by sampling from the Boltzmann distribution over the spin states. However, classical sampling methods such as simulated annealing struggle to efficiently explore these highly frustrated, densely connected energy landscapes, especially when multiple epigenetic markers interact concurrently. Large-scale genomic analyses reveal that nucleosomes with similar epigenetic marks tend to co-fold in the three-dimensional space, suggesting that nucleosome interactions drive chromatin self-organization according to well-defined physical principles, which may be captured by alternative optimization methods.

Quantum annealing (QA) leverages quantum tunneling to escape local minima in rugged energy landscapes, enabling rapid sampling of low-energy, or near-equilibrium, states. Although its precise computational advantage is still being assessed \cite{Albash2018Scaling, nishimori2015comparative}, QA has already proven useful in biologically complex tasks such as protein and RNA folding \cite{PhysRevResearch.ProteinFolding1, PhysRevResearch.ProteinFolding2, wang2024efficient, RNAfolding1, RNAqubo, RNApredicting}, peptide design \cite{peptide, tucs2023quantum}, and conformational transition sampling \cite{ghamari2024sampling}. The same framework can be extended to epigenetic Ising models of chromatin, providing an efficient route to explore TAD configurations. Commercial quantum hardware---e.g., D-Wave’s 5,000-plus-qubit Pegasus processors—can return representative solutions for NP-hard instances in milliseconds, rendering real-time chromatin-state sampling feasible. Via the training process, these machines have shown replicating correlations of a provided data in both supervised and unsupervised settings.\cite{QuantumBoltzmannMachine}

In this work, we embed an Ising-based chromatin model into D-wave quantum processors, parameterized by nucleosome-level epigenetic marks, and benchmark QA against classical simulated annealing.  QA faithfully reproduces the TAD length distributions and insulation score profiles while delivering an order-of-magnitude speedup. These results establish QA as a practical and scalable tool for epigenetic state sampling and a fresh computational lens on genome architecture and its dysregulation.

\label{sec:model}
\section{Sampling and Modeling}

TADs form a hierarchy of nested or overlapping sub-domains. Ongoing chromatin folding brings even distant TADs into proximity, assembling higher-order compartments that stabilize transcriptional programs. Despite this dynamic remodeling, TAD architecture is strikingly conserved through cell divisions and across evolution, underscoring robust organizational principles \cite{ChromatinDomains}.

TADs arise through intertwined physical and biochemical mechanisms—loop extrusion, histone-modifying enzymes, regulatory non-coding RNAs, and architectural proteins such as CTCF and cohesin \cite{ComplexityChromatinFolding, ChromatinFiberPolymer, ChromatinInsulators}. Acting in concert, these factors juxtapose epigenetically similar loci and insulate chromatin domains \cite{ChromatinChromosomeFolding}. Beyond loop extrusion and architectural proteins, emerging evidence highlights the role of biophysical processes such as phase separation in shaping chromatin organization. Multivalent proteins, including transcription factors and coactivators, can form liquid-like condensates through weak, multivalent interactions, creating dynamic hubs that cluster regulatory elements and promote enhancer–promoter communication. These condensates act as an additional physical mechanism contributing to TAD architecture and genome regulation, complementing extrusion-based models \cite{chiariello2020interplay}.

Although mechanistic models have illuminated the molecular basis of TAD organization, data-driven methods—particularly those that use ChIP-seq and Hi-C datasets—excellently describe the boundaries and dynamics of TAD. By correlating chromatin states with spatial organization, such models infer TAD structures and insulation profiles directly from genomic data \cite{IdentificationTopologicalDomains, ChromatinDomains}. 

Entropy-based models \cite{CDFLearning, zhou2016probabilistic, zhou2014global, ansari2019entropy,ansari2015exact,rapp2025} assign probabilistic weights to epigenetic configurations, capturing intermediate TAD states and folding trajectories while linking molecular mechanisms to genome-scale data and paving the way for predictive, targeted epigenetic interventions \cite{CDFLearning}.

Building on this idea, we follow the framework of \cite{CDFLearning}. First, we translate the problem’s energy into a {quadratic unconstrained binary optimization} (QUBO) form.  More precisely, we represent each nucleosome as a binary spin in a QUBO model. Concretely, every binary variable appears at most quadratically in the energy, so the energy can be written as the energy of an Ising model with spin states, local fields,  and pairwise interaction coefficients.  The resulting Ising Hamiltonian, parameterized by epigenomic data, results in the extraction of chromatin domain architectures. Once cast into this QUBO (or equivalently, Ising) format, the problem can be uploaded directly to a quantum-annealing processor, which physically realizes those fields and couplings during the anneal.

Quantum annealing converts the task of sampling from a data probability distribution into repeatedly drawing states from a quantum Ising Hamiltonian whose low-energy manifold encodes that distribution. The resulting problem graph is embedded onto the device topology, with minor-embedding techniques mapping high-degree logical nodes onto chains of physical qubits when necessary. Each anneal begins in a uniform superposition and undergoes a time-dependent evolution in which a strong transverse field is gradually turned off while the problem Hamiltonian is turned on. Quantum tunneling allows the system to traverse tall, narrow energy barriers and settle into low-energy states that follow an approximately Boltzmann distribution $P(s)\propto \exp[-H_{\textup{Ising}}(s)/kT_{\mathrm{eff}}]$, where \(H(s)\) is the QUBO energy of spin configuration \(s\) and \(T_{\mathrm{eff}}\) is the effective sampling temperature.

By collecting thousands of quantum annealing readouts, one obtains an empirical ensemble representing the sampled energy landscape. Post-processing techniques, such as chain-breaking correction through majority voting, gauge averaging, and importance reweighting, are applied to reduce sampling noise and mitigate hardware-induced biases\cite{pazem2023error,pazem2025error}.
Standard statistical tools, including kernel density estimators \cite{silverman2018density}, and variational feedback loops \cite{zhou2022multiple}  are then employed to translate the raw quantum samples into domain-level probability distributions, parameter updates, or predictive observables relevant to the biological context.
 
To accelerate this sampling process and identify low-energy states, we implemented an Ising model on D-Wave quantum annealers. Specifically, (i) chromatin interaction graphs were embedded directly onto the Pegasus hardware topology to minimize chain penalties, and (ii) the resulting quantum samples were benchmarked against experimental datasets to evaluate biological fidelity and sampling efficiency.


\label{sec:Results}
\section{Results}
A detailed description of data preprocessing, model definition, QUBO construction, and embedding procedures is provided in Sec.~IV (Methods).

Below we directly present the sampling results and their interpretation.
All quantum annealing sampling experiments, unless stated otherwise, were performed on the D-Wave Advantage system using the Pegasus $P_{16}$ topology.

We examine the model's ability to sample chromatin states in the vicinity of a template state that exhibits characteristics of a chromatin TAD. Apart from sampling individual chromatin domains, we look at the statistics of sampled chromatin states and compare them with the statistics of the empirical dataset as well as the classical Boltzmann sampling results. In comparison with the classical Boltzmann sampling approach, we examine the allocation and amount of time used for the sampling processes as well as possible optimizations.

\subsection{Sampling of  Empirical Statistics}

\begin{figure}[h!]
\centering
\includegraphics[width=1\columnwidth]{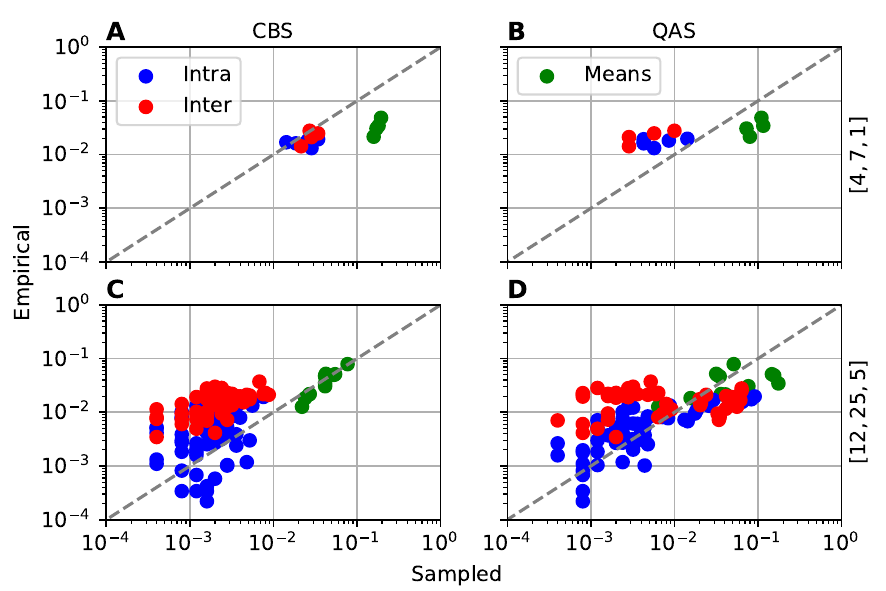}
\caption{
Comparison of empirical and sampled epigenetic incidence statistics. Each column shows results from classical Boltzmann sampling (CBS, left) and Quantum Annealing Sampling (QAS, right). Green points represent mean incidences (diagonal terms), while blue and red points correspond to intra- and inter-nucleosome couplings, respectively.}
\label{fig:results:sampling_stats}
\end{figure}

The main objective of this study is to draw samples from the epigenetic model that replicate the statistical patterns found in the empirical data. Figure~\ref{fig:results:sampling_stats} compares empirical statistics with sampled results, where the left column shows Boltzmann sampling and the right column shows Quantum Annealing Sampling (QAS); green points indicate mean incidences (diagonal terms), while blue and red points correspond to intra- and inter-nucleosome couplings, respectively. Figure \ref{fig:results:sampling_stats} serves as a proof-of-concept: samples generated with quantum annealing reproduce chromatin statistics to relatively the same level of accuracy as those obtained with classical Boltzmann sampling.
This consistency holds across different model sizes, as illustrated by the two examples shown in Figure~\ref{fig:results:sampling_stats}. The statistics are grouped by category, each marked by a distinct color. Within each group, the sampling accuracy is largely self-consistent and can be fine-tuned by adjusting the annealing parameters. However, achieving uniformly high accuracy across all statistical groups simultaneously requires further refinement of the model parameters. This optimization is not pursued in the current study, as the primary focus is to demonstrate qualitative agreement between quantum annealing and classical sampling.

\emph{Annealing Parameters:} For clarity, we briefly summarize the two key embedding-related quantities. The full definitions are provided in Sec.~IV (Methods), including the objective Hamiltonian in Eq.~(\ref{eq:qa:ising_objective}).
Qubit chains are characterized by their length $L_C$, the number of qubits within the chin.
The virtual contraction of qubits within the same logical chain is implemented through the ferromagnetic chain strength $J_C \approx -\max(|J_{ij}|)$, where $J_{ij}$ are the couplings appearing in the effective Ising Hamiltonian. The exact value of $J_C$ is a parameter of the annealing process \cite{DWaveChainStrength} and needs to be adjusted to each individual objective Hamiltonian. In general, there is no fixed rule for appropriately setting $J_C$; rather, it requires experimentation and parameter optimization.

A grid search over the annealing time \(T_{A}\) and the ferromagnetic chain strength $J_C$  was carried out to determine optimal sampling conditions.  To find the optimal annealing parameters, we swept a range of annealing times $T_A$ and chain strengths $J_C$ and evaluated the resulting sampled statistics using $R^2$ to determine the annealing parameters that yield the best performance.
We summarize the metric definitions here for clarity;  details are provided in Sec.~IV (Methods).
$R^2$ is the coefficient of determination of the quality line between empirical and sampling statistics and is defined as $R^{2}=1- {\sum_{i}\bigl(\rho^{\text{emp}}_{i}-\rho^{\text{sample}}_{i}\bigr)^{2}} / {\sum_{i}\bigl(\rho^{\text{emp}}_{i}-\overline{\rho}^{\text{emp}}\bigr)^{2}}$, with the empirical and sampling statistics for each observable $i$ being $\rho^\text{emp}$ and $\rho^\text{sample}$, respectively, and \(\overline{\rho}^{\text{emp}}\) is their empirical mean value. Because individual statistics span several orders of magnitude, $R^2$ is calculated for the logarithm of each statistical value. Our search for optimization is based on maximizing $R^2$. 
Tests across different sampling settings and optimization sweeps show that 
\(R^{2}\) exhibits a single broad optimum as a function of the chain strength 
\(J_{C}\). If \(J_{C}\) is \emph{too small},  the ferromagnetic coupling that binds the physical qubits in a chain is not strong enough to keep them aligned. This results in chain breaks. 

Moreover, the annealing time \(T_{A}\), unlike optimization tasks (which favor ever-longer sweeps), shows an \emph{early} optimum: if \(T_{A}\) is prolonged further, the system relaxes almost exclusively into the ground state and fails to explore the full Boltzmann distribution. The relative amount of chain breaks in these sampling configurations is approximately $10\,\%$.

\begin{figure}[htbp]
  \centering
  \includegraphics[width=\columnwidth]{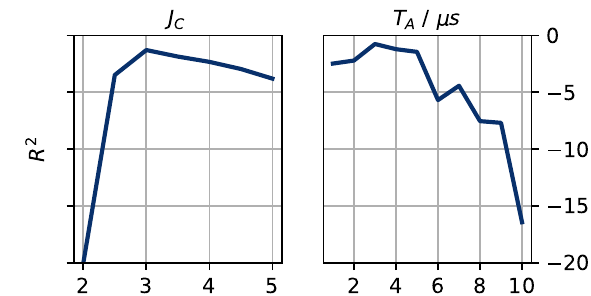}
  \caption{ Grid search of annealing time \(T_{A}\) (Left) and chain strength \(J_{C}\) (Right).  The y-axis indicates \(R^{2}\).   The x axis represents the unitless chain strength $J_C$ (left) and the annealing time (right). The chain strength shows a typical global optimum, while the annealing time shows an early optimum.}
  \label{fig:results:annealing_parameters}
\end{figure}

A grid search over the annealing time $T_A$ and chain strength $J_C$ reveals optimal parameter regions that exhibit characteristic behavior unique to sampling applications—distinct from optimization use cases. These optimal settings enable more accurate generation of chromatin states under QAS and highlight the importance of tuning quantum schedules for biological sampling tasks.

\begin{figure}[htbp]
  \centering
  \includegraphics[width=1.0\columnwidth]{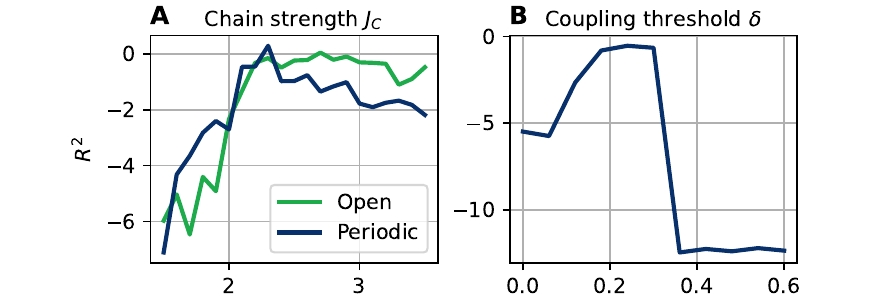}
  \caption{(A)~Performance comparison for open vs.\ periodic boundary
  conditions as a function of \(J_{C}\).
  (B)~Model performance as weak logical couplings below a threshold
  \(\delta\) are pruned before embedding.}
  \label{fig:results:boundary_conditions_and_coupling_threshold}
\end{figure}

\emph {Boundary condition effect:}  We consider a hardware graph and map onto it an objective graph with either periodic or open boundary conditions.  The results of such sampling have been plotted in Figure \ref{fig:results:boundary_conditions_and_coupling_threshold}A, which shows a comparison between the optimum statistics.  Results show that both boundary conditions are advantageous in different regions of the chain strength $J_C$. $J_C$ refers to a physical ZZ-coupling between two physical qubits in a logical qubit.

For small \(J_{C}\), periodic embeddings perform better because they treat every 
nucleosome equivalently. However, for large \(J_{C}\), open embeddings become 
advantageous, as they map to shorter physical chains and are therefore easier to 
embed.

It is interesting that by discarding the logical couplings with \(|J|<\delta\) and treating the incidences as non-interacting, we can simplify the embedding. This simplification can \emph{improve} performance up to \(\delta\approx 0.25\)  without
removing biologically relevant interactions.  Beyond that point crucial interactions (e.g.\ markers H3K4me1–H3K4me2) are lost and \(R^{2}\) plummets. Therefore, it is useful to consider that \(\delta\) is a useful knob, but only within a model-dependent
  safe window.  Panel~B shows the effect of pruning logical couplings whose strength is below \(\delta\).

The take home message from sampling empirical statistics in D-Wave's quantum annealer is the following:  careful tuning of \(T_{A}\), \(J_{C}\), boundary conditions, and the coupling threshold \(\delta\) can bring quantum annealer samples into close quantitative agreement with empirical chromatin statistics \emph{without} altering the underlying biophysical parameters.

\begin{figure}[h!]
\centering
\includegraphics[width=\columnwidth]{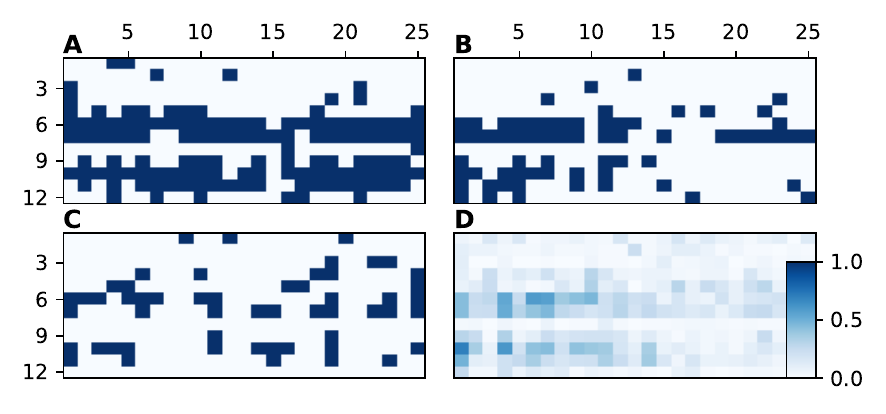}
\caption{Three representative unbiased samples drawn directly from a quantum annealing process (A-C) and the mean distribution of incidences during that sampling (D). These samples are shown only to illustrate the qualitative appearance of raw QA outputs. X-axis shows indices of epigenetic markers and Y-axis shows nucleosome indices. Samples were taken with $T_A = 20 \, \mu s$ and $J_C=2$.}
\label{fig:results:free_samples}
\end{figure}

\emph{Examples:} Fig.\ref{fig:results:free_samples} illustrates what quantum annealing samples look like. These visualizations serve only to show the qualitative form of the sampler’s output. Figure \ref{fig:results:free_samples} shows a number of different samples and the mean incidence distribution of $100$ samples. It is apparent that selected samples exhibit similarity to chromatin TADs and display strong correlations between certain epigenetic markers (e.g. between H3K4me1, H3K4me2 and H3K27ac, see  Supplementary Material \cite{kempe2025}. Other states reveal no TAD-like structures, but rather single incidences; this is also typical for the empirical data and contributes to reproducing given statistics, but yields no benefit for chromatin domain predictions. In figure \ref{fig:results:free_samples}D, the mean occurrence per incidence is depicted for $100$ samples.

\subsection{Sampling of Empirical Domains}

We next assess how well the annealer can reproduce \emph{exact}
empirical incidence states  \emph{and} states that lie in their structural vicinity.

\emph{Structural (Hamming) distance}: Let the empirical incidence pattern be 
\( \mathbf{A}=(A_{1},A_{2},\dots ,A_{n})^{\mathsf T} \). Each entry is an empirical incidence that is obtained directly from the biological data. $A_i$ is $+1$ if a marker is present and $-1$ if not.

Similarly we define $x_i$ to be corresponding incidence proposed by the
quantum-annealer sample. Therefore the analoge of $A$ reproduced on a quantum annealer is \( \mathbf{x}=(x_{1},x_{2},\dots ,x_{n})^{\mathsf T}
\). Every entry \( A_{i},x_{i}\in\{+1,-1\} \) answers a yes/no question such as  ``is marker \emph{m} present on nucleosome \emph{n}?''. We summarize the metric definitions here for clarity; more details are provided in Sec.~IV (Methods). The \emph{structural (Hamming) distance} between the two patterns is
\begin{equation}
  D_{\mathbf{A}}(\mathbf{x})
  =\frac12\sum_{i=1}^{n}\bigl(1-A_{i}\,x_{i}\bigr).
  \label{eq:hamming}
\end{equation}

The distance can be interpreted in the following manner:         For each index \(i\) we have \(A_{i}x_{i}=+1\) if the two patterns agree and \(-1\) if they disagree.  Consequently
 \(1-A_{i}x_{i}\) equals \(0\) (agreement) or \(2\) (disagreement). To compare models of $M_A$ markers and $N_A$ nucleosomes, we normalize~\eqref{eq:hamming}:
\begin{equation}
  d_{\mathbf{A}}(\mathbf{x})
  \;=\;
  \frac{D_{\mathbf{A}}(\mathbf{x})}{M_A N_A}
  \;\in\;[0,1],
  \label{eq:relative_hamming}
\end{equation}
where \(d_{\mathbf{A}}=0\) indicates perfect match with the empirical state and \(d_{\mathbf{A}}=1\) indicates every incidence is flipped. Both \(D_{\mathbf{A}}\) and \(d_{\mathbf{A}}\) quantify how ``close'' a sampled configuration is to the true biological pattern, without requiring the sampler to \textit{exactly} recreate a single ground state.

 \begin{figure}[h]
\centering
\includegraphics[width=\columnwidth]{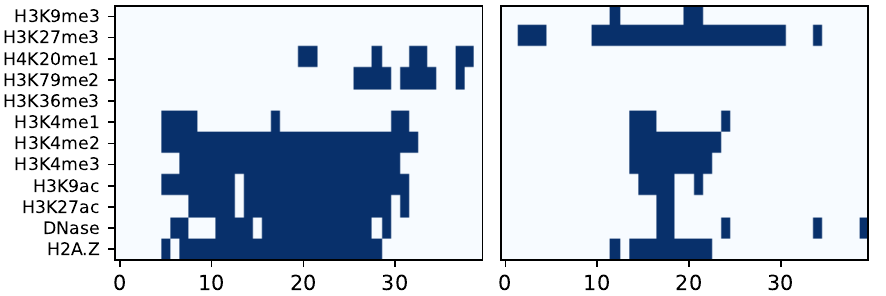}
\caption{Example of a promoter domain (left) and a bivalent domain (right) in an empirical dataset in binarized form. The promoter domain is marked by the presence of the H3K4me3 mark, while the bivalent domain is shown by the combination of the H3K27me3 mark and any of the H3K4 methylations. Activations are marked blue. X axis shows nucleosome position; Y axis shows nucleosomic marks.}
\label{fig:bio:example_domains}
\end{figure}

\begin{figure}[h!]
\centering
\includegraphics[width=1.1\columnwidth]{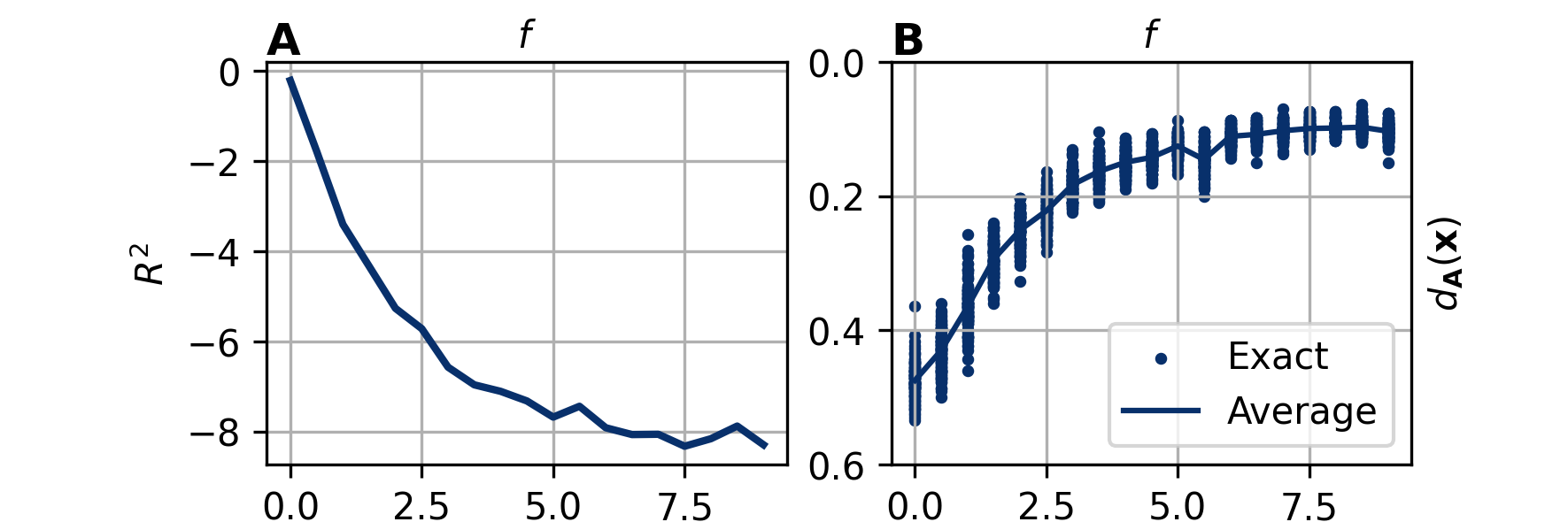}
\caption{Effect of the bias strength $f$ on the model's sampling performance and the similarity between sampling states and empirical state. (A) Decreasing performance index $R^2$ for increasing bias strength. (B) Exact and average values of relative structural distance between sampled and empirical states for rising bias strength.}
\label{fig:results:incidence_bias_performance}
\end{figure}

\emph{Annealing parameters}: Using the incidence bias model \cite{kempethesis}, the effect of the bias strength $f$ is examined with incidence biases taken from a data section of a promoter TAD. 
 Here, the bias strength $f$ scales the local fields associated with the empirical template state $\mathbf{x}_A$, effectively adding a linear penalty that encourages the sampled configurations to align with the empirical pattern without modifying the pairwise couplings of the model.
In contrast to Figure~\ref{fig:results:free_samples}, Figure~\ref{fig:bio:example_domains} depicts empirical incidence patterns. These serve as reference structures (e.g., promoter and bivalent domains) and illustrate the chromatin motifs we ultimately compare against. As visible in figure \ref{fig:results:incidence_bias_performance}, the model performance declines for growing bias strength, since the template state $\mathbf{x}_A$ does not represent overall statistics. 

The structural difference, however shrinks with growing $f$ which shows that the vicinity of the template state can be sampled with controlling parameter $f$. These benchmarks use a standard annealing schedule as presented in Supplementary Material \cite{kempe2025}.

\begin{figure}[h!]
\centering
\includegraphics[width=\columnwidth]{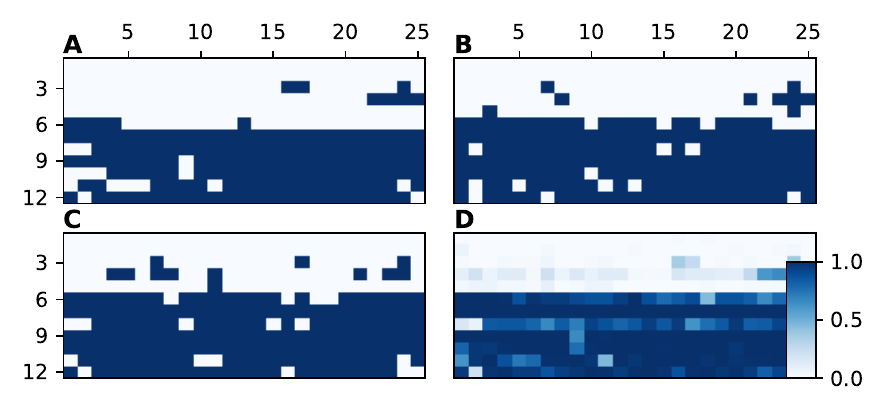}
\caption{(A)~Empirical chromatin state from a promoter domain.
         (B,C)~Sampled states at bias strength \(f=5\).
         (D)~Average incidence map over 100 samples at \(f=5\).
         Structural motifs of (A) are still discernible in all cases.
         X axis shows nucleosome position; Y axis shows nucleosomic marks (numbered 1-12).}
\label{fig:results:biased_samples}
\end{figure}

\emph{Examples}: Figure~\ref{fig:results:biased_samples} illustrates representative samples obtained at a bias strength of \(f = 5\), along with their associated energy levels and structural differences from the empirical template state. Panels~(B) and~(C) show two individual sampled states, while panel~(D) displays the average incidence map over 100 such samples. In all cases, key structural features of the empirical state (panel~A) remain visible, demonstrating that the bias effectively confines sampling to a meaningful vicinity around the biological reference. Figure~\ref{fig:results:biased_samples} connects the previous figures by demonstrating how the sampler behaves when explicitly biased toward an empirical TAD configuration from Figure~\ref{fig:bio:example_domains}. In this setting, the QA samples cluster in the structural vicinity of the empirical state, unlike the unbiased samples shown in Figure~\ref{fig:results:free_samples}.

\emph{Reverse annealing}: 
As an alternative to introducing bias directly into the model parameters, it is also possible to guide the sampling process by initializing the qubits in a predefined configuration—specifically, the empirical state—and applying a \emph{reverse annealing} schedule~\cite{ReverseQuantumAnnealing}. This technique preserves the original energy landscape: no changes are made to the couplings or local fields. Instead, the system begins in a classical state aligned with the empirical template and is briefly pushed out of equilibrium before being re-annealed.

The reverse annealing protocol begins at \(s = 1\), where the system is initialized in a classical state. 
It then reverses the annealing schedule to a minimum point \(s = s_{R}\) at time \(t_{R}\), partially reintroducing quantum fluctuations, before linearly returning to \(s = 1\).

\begin{figure}[h!]
\centering
\includegraphics[width=\columnwidth]{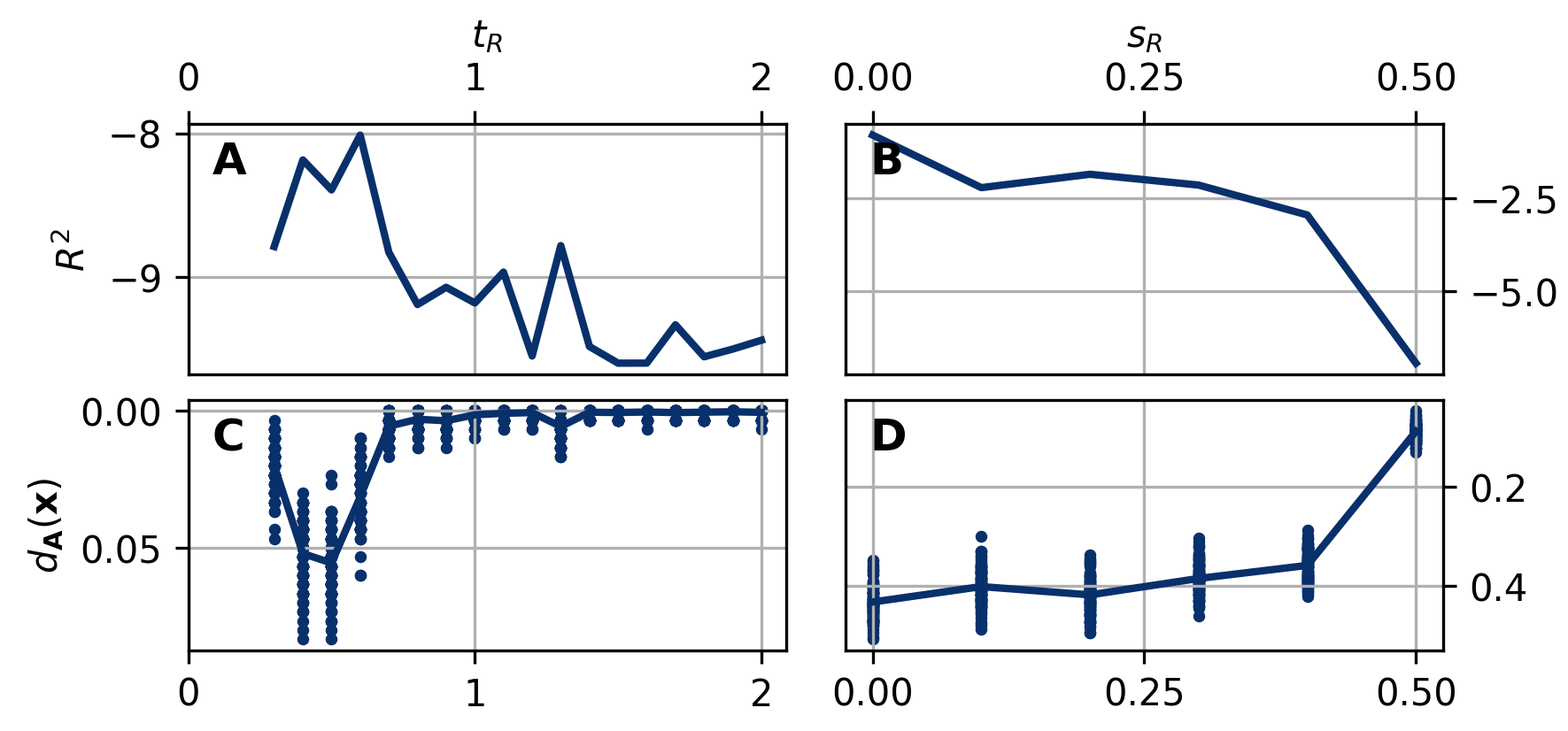}
\caption{Sampling performance using reverse annealing.
Top: \(R^2\) accuracy relative to empirical statistics.
Bottom: Relative structural distance \(d_{\mathbf{A}}(\mathbf{x})\).
(Left) Varying return time \(t_{R}\) at fixed depth \(s_{R} = 0.4\).
(Right) Varying reversing depth \(s_{R}\) at fixed return time \(t_{R} = 200\,\mathrm{ns}\).}
\label{fig:results:reverse_annealing_performance}
\end{figure}

This allows partial exploration around the initial state without fully resetting the system. As shown in Figure~\ref{fig:results:reverse_annealing_performance}, both the return time \(t_{R}\) and the reversing depth \(s_{R}\) critically affect the sampling behavior.

\begin{itemize}
  \item \textbf{Return time \(t_{R}\)} (left column): Increasing \(t_{R}\) keeps the system longer in the reversed portion of the annealing schedule, improving sampling quality (higher \(R^{2}\)) but also increasing the structural distance \(d_{\mathbf{A}}(\mathbf{x})\).
  \item \textbf{Reversing depth \(s_{R}\)} (right column): Shallower reversals (\(s_{R} \to 1\)) limit how far the schedule moves away from the classical state, keeping samples close to the template (low structural distance) but reducing variability and hence lowering \(R^{2}\).
\end{itemize}

Overall, reverse annealing offers a controlled way to bias the sampling toward empirical configurations while maintaining the integrity of the trained model parameters. It is especially useful when seeking to sample in the local neighborhood of a known reference state.

\subsection{Execution Time}
\label{sec:execution_time}

A direct comparison of time complexity between classical Boltzmann sampling and quantum annealing sampling is inherently challenging due to the fundamentally different computational paradigms they operate under: classical sampling involves sequential, algorithmic transitions in a gate-based architecture, whereas quantum annealing relies on a continuously evolving physical system. Moreover, since the quantum sampling process does not necessarily follow an adiabatic path, the adiabatic theorem does not provide a relevant theoretical bound on runtime.

\begin{figure}[t]
\centering
\includegraphics[width=\columnwidth]{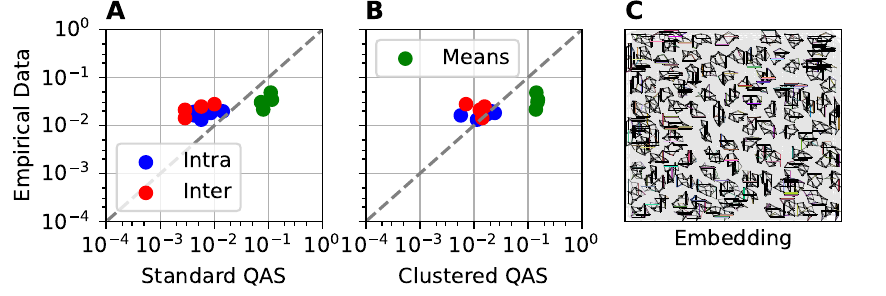}
\caption{Cluster-based parallelization of quantum sampling.
(A)~Statistics obtained from standard sequential sampling.
(B)~Statistics obtained from a single anneal using a replicated cluster graph, where multiple independent sampling Hamiltonians are embedded and sampled concurrently on the same quantum processing unit.
(C)~Embedding of the cluster graph on the Pegasus chip topology. Panels~(A) and~(B) exhibit comparable statistics for the same model.
Using the mean absolute log-space distance to the empirical statistics, the standard and clustered workflows yield values of $0.51$ and $0.31$.
This figure shows workflow equivalence between sequential and clustered quantum sampling and validates that cluster replication does not degrade sampling quality.}
\label{fig:results:cluster_parallelization}
\end{figure}

To make a meaningful comparison, we evaluate the wall-clock execution time by contrasting the time allocations of major process steps in each method. For consistency, we assume an equal number of samples \(n_{\text{sample}} = 100\) for both classical and quantum approaches, as this is the baseline for computing model statistics.
To avoid overstating our claims, we emphasize that our timing discussion is qualitative and at the workflow level. We have focused on hardware level parallelism and amortization of embedding costs (Figure~\ref{fig:results:cluster_parallelization}).
This amortization refers to a practical reduction in average per sample overhead and does not imply an asymptotic or algorithmic scaling advantage.

\emph{Parallelism and speed-up}: Both methods can, in principle, benefit from parallel execution. In classical Boltzmann sampling, parallelization is achieved through multiprocessing—e.g., using Python's \texttt{multiprocessing} API—which yields a speed-up of approximately \(4.13\times\) relative to serial execution for this sample size.

Quantum annealing enables a different kind of parallelism: by replicating the objective graph into \(n_{\text{sample}}\) independent subgraphs (forming a cluster graph), one can obtain all samples in a single quantum readout. This is possible only if the quantum chip is large enough to accommodate all subgraphs simultaneously. For instance, a model of size \([4,7,1]\) can be duplicated 100 times and embedded on the Pegasus topology.

Figure~\ref{fig:results:cluster_parallelization} compares the results of such cluster-based parallelization to the standard sequential quantum sampling.

As panels (A) and (B) show, cluster-based parallelization yields statistics that are comparable to those obtained from standard sequential quantum annealing sampling. Using the mean absolute log-space distance to the empirical statistics, 
we observe that cluster replication produces errors of comparable size in the same order of magnitude range to those obtained with sequential sampling, while enabling single-anneal throughput.
Because all 100 samples are collected in one quantum anneal, the speed-up, based solely on annealing time, is effectively \(100\times\).
We emphasize that Figure~\ref{fig:results:cluster_parallelization}  is intended to assess whether cluster-based parallelization alters the statistical behavior of the quantum sampler, while Figure~\ref {fig:results:sampling_stats} evaluates the biological accuracy

\emph{Process comparison}:  In classical Boltzmann sampling, each sample is generated via a Markov chain that evolves step-by-step from a random initial state. At every step, a random incidence variable is selected, and the corresponding energy difference must be computed to accept or reject the new state. As each step depends on the previous one, this process is inherently sequential. The total time is thus proportional to the number of sampling steps \(n_{\text{steps}}\) per sample.

Quantum annealing, in contrast, generates each sample in three stages: (1) qubit initialization, (2) physical annealing, and (3) measurement. This process is non-iterative per sample. However, before sampling can begin, a suitable embedding of the model onto the hardware graph must be computed. For large problem sizes, this embedding process can dominate the overall runtime. Crucially though, the embedding only depends on the model size \([M,N,L]\) and not on the specific parameter values. Once computed, the same embedding can be reused across all samples with identical model structure.

\medskip
In summary, while quantum annealing offers significant speed-ups through hardware-level parallelism, especially when cluster replication is feasible, classical methods remain competitive in flexibility and scalability—especially when the quantum embedding overhead becomes non-negligible.

\label{sec:Method}
\section{Methods}
This section provides the methodological details underlying the results presented in Sec.~III. It includes the preprocessing steps, model formulation, QUBO objectives, learning, and quantum‑annealing embedding procedures.

In this section, we first discuss in Sec.~(\ref{sec. sampw}) how we sample from the statistical epigenetic model using QA, which requires outlining the steps used for both classical and quantum sampling, including binarization, the relevant statistics, and the model definition. This is followed by the specific steps for quantum-annealing sampling. In Secs.~(\ref{sec. qubo},\ref{sex.cartesian},\ref{sec. learning}), we describe the QUBO objectives, Cartesian modeling, and parameter learning, respectively. Finally, in Sec.~(\ref{sec:dwave_embedding}), we present the embedding of the epigenetic problem on a quantum annealer and discuss the embedding criteria considered.    

\subsection{Sampling workflow}
\label{sec. sampw}
In order to sample from the statistical epigenetic model using QA, the statistics first need to be extracted from the ChIP-based dataset and the model needs to be adapted to a QA device. The steps associated with this preparation are listed below. 
In Sec.~\ref{sec:Results} we present the sampling results; here we describe the methodological steps required to obtain them.

Firstly, we discuss the steps required for both classical and quantum sampling.

\begin{description}
    \item[Binarization] Since the model is based on binary incidences of epigenetic markers on different nucleosome locations, we first binarize and---for better computational efficiency---bin the (semi-)continuous format of epigenetic markers in \cite{RoadmapEpigenomicsProject, IMR90}. In the chromatin structure, each nucleosome corresponds to a 200 base-pair section of DNA.  We used data from the IMR90 fetal lung fibroblast cell line provided by the NIH Roadmap Epigenomics Project \cite{RoadmapEpigenomicsProject, IMR90}. This dataset includes 12 epigenetic marks measured across approximately 140 million base pairs on chromosome 9. The dataset is divided into 200 bp bins, assuming one nucleosome per bin. An epigenetic marker is labeled as active if its peak strength surpasses a threshold, resulting in 700,000 bins and 8.4 million binary incidences across the 12 markers. Each bin can exhibit one of 4,096 possible combinations of markers.

    \item[Statistics] We compute statistical features of binarized empirical data as a reference for the learning procedure and later sampling results. The binarized data is represented as a binary matrix $A$, where each element indicates the presence (1) or absence (0) of a marker at a specific nucleosome position. The individual incidence cells are addressed by their epigenetic marker $ m \in [0, M - 1] $ and the nucleosome position $ n \in [0, N - 1] $, with the number of markers and nucleosomes being $M$ and $N$. The binarized dataset can then be denoted as the matrix $\mathbf{x}$ with elements: $x_m^n \in \{0, 1\}$. It is important to distinguish between the empirical matrix $\mathbf{A}$ and the model variable $\mathbf{x}$. Both share the same binary structure, where entries indicate the presence or absence of specific epigenetic marks at given nucleosome positions. However, $\mathbf{A}$ is fixed and derived from experimental data, serving as the reference for statistical constraints. In contrast, $\mathbf{x}$ represents a configuration generated by the model during sampling or optimization. Thus, $\mathbf{A}$ encodes observed incidences, while $\mathbf{x}$ explores possible states and can evolve based on the model.

    Key statistics, such as the mean incidence of markers  
    $\mu_m = \frac{1}{N} \sum_n x_m^n$ and correlations between markers within $\rho_{mm'} = \frac{1}{N} \sum_n x_m^n \cdot x_{m'}^n$ and across nucleosomes $\rho_m^l = \frac{1}{N} \sum_n x_m^n\cdot x_m^{n+l}$,  are computed to describe the dataset's distribution and are later used to evaluate sampled chromatin states.
    
    \item[Model] We define a model based on epigenetic marker incidences that maps probable chromatin states according to the recorded statistics to low-energy values of an objective function. We parameterize the model to accurately produce the given statistics during sampling.
\end{description}

Until this point, the steps described are common to both classical and quantum sampling: they define the model, specify the variables and statistics of interest, and establish the discretization/binarization and objective formulation needed to generate samples in a controlled and reproducible way. In other words, these steps set up a unified sampling problem that can be addressed with either a classical sampler or a quantum device.

We now turn to the steps that are specific to quantum-annealing sampling. These focus on mapping the prepared objective to the annealer’s native constraints, selecting an appropriate embedding onto the hardware graph, and specifying annealing and readout settings (e.g., chain strength, anneal schedule, and post-processing) required to obtain and interpret samples from the quantum processor.

\begin{description}
\setcounter{enumi}{3}
\item[Objective] 
We define our chromatin model slightly differently from the classical Ising model, which consists of a one-dimensional chain of spins with nearest-neighbor interactions. Our chromatin model is represented by a two-dimensional binary tensor~\(\mathbf{x}\). Rows correspond to nucleosome positions and columns to epigenetic markers, so each variable \(x_{m}^{\,n}\in\{0,1\}\) encodes whether marker \(m\) (e.g., H3K27ac, CpG-methylation) is present (\(1\)) or absent (\(0\)) at nucleosome position \(n\). This tensor is a classical data structure used to define the QUBO objective function. This structure introduces two orthogonal interaction dimensions: (i) intra-nucleosome couplings between different markers at the same position, and (ii) inter-nucleosome couplings between identical markers across positions. Therefore, the resulting interaction graph is not linear, reflecting the complexity of chromatin organization. We define an objective function in a consistent way with the model encoded in the energy expectation of the quantum system's Hamiltonian: 
\begin{equation}
H(\mathbf{x}) = \sum_{m \geq m', n \geq n'} Q_{mm'}^{nn'} x_m^n x_{m'}^{n'},
\end{equation}
where conditions $m>m'$ and $n>n'$ are to avoid repeated counting of the same state.  This reflects the fact that the second qubit represents markers and locations on one side of the first qubit. This ensures defining the couplings \( \mathbf{Q} \) in a unique manner and restricting the lower triangle of a qubit adjacency matrix.

\item[Embedding] We need to embed the objective Hamiltonian onto the quantum processor's specific topology in a D-Wave quantum annealing device.

\item[Sampling] Using the parameters computed in step 3, we sample the states and anneal multiple times with the quantum annealer.

\item[Analysis] We compare and analyze statistical distributions from quantum‑annealer samples with empirical chromatin‑state datasets.
\end{description}

\subsection{QUBO Objectives}
\label{sec. qubo}

As discussed above, our binary incidence tensor  
\(x_{m}^{\,n}\in\{0,1\}\) records whether epigenetic marker 
\(m\)  is present (\(1\)) or absent (\(0\)) at
nucleosome position \(n\). 
When this tensor is assigned to an unconstrained quadratic binary optimization or, in summary, QUBO, every binary variable is assigned (i) a \emph{bias}, the energy cost or preference to flip that single bit, and
(ii) \emph{couplings} to other bits that capture pairwise correlations.

{\bf The single variable bias matrix} \(q\) contains entries of the diagonal
elements of the QUBO matrix \(Q\), that is, $ q_{m}^{\,n} \;\equiv \; Q_{m,m}^{n,n} $, therefore \(q_{m}^{\,n}\) quantifies the energetic tendency for marker \(m\) to be
present in nucleosome \(n\). The statistical models assume equal statistical mean and correlation for each nucleosome and thereby share the linear coupling strengths predicted by the mean as $\forall n$, ~$n' \in [0 \dots N - 1]$, ~i.e ~$q_m^n = q_m^{n'}$.

\begin{figure}[h]
\centering
\includegraphics[width=\columnwidth]{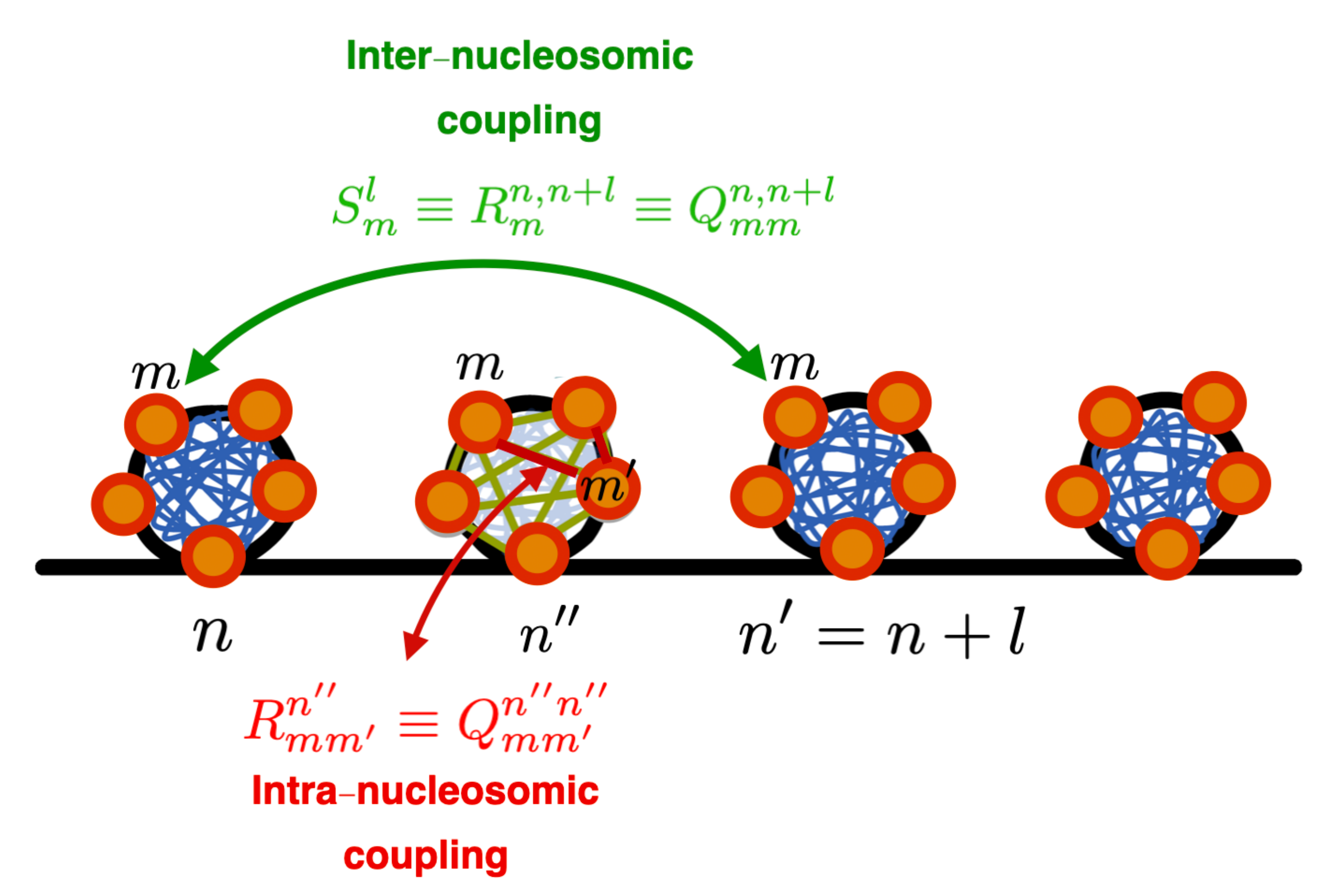}
\caption{Sketch of the binarized chromatin model consisting of four nucleosomes. Epigenetic markers $m$ and $m'$ on the same nucleosome $n''$ interact by the intra-nucleosomic coupling $R_{mm'}^{n''}$ shown in red, while the same epigenetic marker $m$ on two different nucleosomes $n, n'=n+l$ interact by the inter-nucleosome coupling $S_m^l$, shown in green.}
\label{fig:model:model_sketch}
\end{figure}

{\bf Intra-nucleosome couplings} are  coupling strengths between different epigenetic markers $m$ and $m'$ at the same nucleosome locations $n$, denoted by dropping a nuclesome's index from $Q$; i.e. $R_{mm'}^{\,n} \;\equiv \; Q_{m,m'}^{n,n}$ with $ m\neq m'$ for $\forall n$,  ~$n' \in [0 \dots N - 1]$ (See Figure \ref{fig:model:model_sketch}).  These terms favour—or penalise—the simultaneous appearance of markers \(m\) and
\(m'\) at position \(n\), reflecting biochemical cross-talk within a single nucleosome location. 

The statistical models assume the coupling of any pair of nucleosomes $m, m'$ at any location $n$ (or $n'$) are the same, i.e.  $R_{mm'}^n = R_{mm'}^{n'}$, namely $Q_{mm'}^{nn}=Q_{mm'}^{n'n'}$ for all $\forall n, n' \in [0 \dots N - 1]$.

{\bf Inter-nucleosome couplings}
 are couplings between a single epigenetic marker $m$ at two different nucleosomes $n$ and $n'$, regarded as ``inter-nucleosomic'' couplings. In general, these can be further simplified to  $R_{m}^{n n'} = Q_{mm'}^{n n'}$ for $\forall m,m'$,  ~$m,m' \in [0 \dots M - 1]$  (See Figure \ref{fig:model:model_sketch}).  

 In most cases, the coupling between variables of different nucleosome positions primarily depends on the distance between the two nucleosomes, not on the label of nucleosome location. For that reason, it is sensible to define inter-nucleosomic couplings using the nucleosome distance \( l = n' - n \) with \( l \in [1 \dots L] \) where \( L \in \mathbb{N}^+ \) is the inter-nucleosomic coupling distance considered in the model, where the maximum is $N$, therefore $R_{m}^{n, n+l} = R_{m'}^{n', n'+l} $ for $\forall n, n' \in [0 \dots N - l - 1]$. Considering further simplification that the coupling is the same in all nucleosome positions, this coupling can be easily expressed by two indices, marker $m$ and the location $l$, i.e.  $S_m^l \equiv R_m^{n, n+l}$ for $\forall n \in [0 \dots N-l-1]$.

{\bf Periodic boundary condition} helps to mitigate boundary effects for the couplings concerning two different nucleosome sites. As an example, periodic inter-nucleosomic couplings would result in the equality $S_m^l = Q_{mm}^{n, n+l\bmod{N}}$  if $Q_{mm}^{n, n+l} = Q_{m m}^{n', n'+l}$ for $\forall n, n' \in [0 \dots N - l - 1]$. While improving overall statistics and the generality of the model, introducing periodic boundary conditions will result in a more complex embedding. We will discuss this in section \ref{sec:dwave_embedding}.  

{\bf Other mixed couplings} between incidences of different markers at different nucleosome sites do not bear significant statistical relevance and would complicate the computational model non-trivially. Therefore, these couplings do not need to be included in the objective graph; they have an effective weight of zero. Consequently, the following coupling elements will be zero for all considered models:
\begin{equation}
 Q_{mm'}^{nn'} = 0 \quad \forall m \neq m', \quad n \neq n'
\label{eq. zero Q}   
\end{equation}

Including near-zero couplings results in a significantly more complex embedding and can hamper the performance of the model. Therefore, a correlation coupling threshold \( \delta \) is introduced. All couplings below this value are removed entirely from the objective graph, simplifying both the embedding process and the final annealing graph.

\subsection{The ``Cartesian'' modeling}
\label{sex.cartesian}
Consider that we represent $M$ markers on a vertical line of dots, where nucleosome location changes horizontaly. In this representation, the exclusion of mixed couplings in Eq. (\ref{eq. zero Q}) effectively makes the model structurally ``Cartesian'', since only  ``horizontal'' (inter-nucleosome) and ``vertical'' (intra-nucleosome) couplings are nonzero. 

Consequently, the effective objective function is
\begin{equation}
\begin{aligned}
H(\mathbf{x}) = \sum_{m=0}^{M-1}\sum_{n=0}^{N-1} q_m^n x_m^n + \sum^{M-1}_{\substack{m_1,m_2=0 \\ m_1>m_2}} \sum_{n=0}^{N-1} R_{m_1 m_2}^n x_{m_1}^n x_{m_2}^n  \\
 + \sum_{m=0}^{M-1}  \sum_{l=1}^{N}  \sum_{\substack{n=0 \\ n+l \bmod N}}^{N-1} S_m^l x_m^n x_m^{n+l}
 \end{aligned}
 \label{eq.Hx}
\end{equation}
where all the bias and coupling parameters will be set differently depending on the actual model.

An extra weighting between the different types of parameters is realized implicitly via the size of the individual parameters. Apart from the Ising parameters, the main structural parameters that describe this model can be summarized as:
\begin{itemize}
    \item number of epigenetic markers $M$
    \item number of nucleosomes $N$
    \item maximum correlation length $L$,
\end{itemize}
which is abbreviated as $[M,N,L]$. Considering the annealing process, additional hyper-parameters are annealing time $T_A$, chain strength $J_C$ and number of samples $n_{smpl}$. Adhering to suggestions made in \cite{CDFLearning} based on biochemical importance, the maximum model size is set to $M_\text{max} = 12$, $N_\text{max} = 25$ and $L_\text{max} = 5$, i.e. $[12, 25, 5]$. Due to the nucleosomically shared QUBO parameters that produce a nucleosome-independent objective\footnote{(meaning that states that only differ by a periodic shift in nucleosome positions produce the same objective energy)}, the model can be regarded as ``convolutional'' in analogy to positionally shared weights in convolutional neural networks \cite{lecunCNN, pinaya2020convolutional}.

The Hamiltonian in Eq.~(3) defines an energy landscape over binary incidence states of epigenetic marks along the chromatin fiber. A \emph{chromatin configuration} refers to one such state, represented by the tensor $\mathbf{x}$, which encodes the presence or absence of specific marks at each nucleosome position. While $\mathbf{x}$ can be observed experimentally for real cells, our goal is to \emph{sample} configurations from the model that reproduce the statistical properties of empirical data rather than minimize to a single ground state. The coupling parameters $Q$ in Eq.~(3) are optimized during the training phase using experimental statistics (Section~II.4), ensuring that low-energy states correspond to biologically plausible domain structures. Thus, optimization applies to $Q$, whereas sampling explores ensembles of $\mathbf{x}$ configurations consistent with the learned couplings. The main goal of these models is to sample chromatin configurations that resemble the structural features of the real chromatin domains observed in the experimental genomic data. To achieve this, the models are designed to follow the global statistical properties of the data while being energetically biased toward forming structures similar—but not identical—to those empirically observed.

This allows for tuning between two conceptual extremes:  1) Highly specific models: One can design a model that captures local biases and correlations directly from the empirical data, effectively mapping each variable in the model to a specific genomic location. In this case, quantum sampling yields very few states that differ from the empirical configuration, making it unsuitable for generating diverse alternatives. Interestingly, the same Ising model, when sampled with classical simulated annealing, performs significantly worse in terms of success rate per time.   2)	Fully generic models: Alternatively, one can construct models based solely on global statistics, without introducing any location-specific biases. While this approach is maximally general, it struggles to generate TAD structures that meaningfully resemble those found in the empirical data. Between these two extremes lie a range of intermediate strategies, which are described below and analyzed in detail in Section \ref{sec:Results}.

\
\subsection{Learning Parameters}
\label{sec. learning}
To reproduce the statistical properties of the empirical dataset, suitable model parameters must be learned. To this end, we employ a classical optimization procedure that iteratively updates the model parameters using stochastic gradient descent, guided by samples generated via a classical Boltzmann sampling algorithm. This learning scheme is inspired by the approach introduced in \cite{CDFLearning}, although notable differences exist in the specific optimization steps.

\begin{figure}[t]
\centering
\includegraphics[width=\columnwidth]{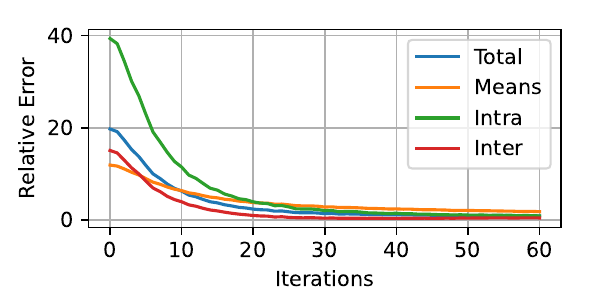}
\caption{Learning progression of parameter optimization using iterative classical Boltzmann sampling. The error is the difference between the model statistics and the empirical data statistics for the mean incidence markers, the correlations between markers within and across nucleosomes, and the sum of the mean, intranucleosome, and cross-nucleosome values.}
\label{fig:model:errors}
\end{figure}

The parameter learning procedure consists of the following steps:
\begin{enumerate}
\item  Initialize the model with random parameters. Starting from a random point in parameter space ensures exploration of diverse regions in the optimization landscape.
\item  Generate multiple samples from the model using Boltzmann sampling. Sampling generates configurations according to the current model's energy landscape, mimicking physical equilibrium behavior.
\item  Compute statistical observables from the generated samples. Quantities such as marginal probabilities and pairwise correlations are calculated to compare with empirical data.
\item  Adjust the model parameters according to the discrepancy between simulated and empirical statistics. The difference between the observed and simulated statistics defines a loss function, minimized by updating the parameters.
\item  Repeat steps 2–4 until the total error falls below a threshold. The iteration continues until the model statistically reproduces key features of the empirical dataset.
\end{enumerate}

The Boltzmann Sampling works with the following steps: 

\begin{enumerate}
\item Start with a random model state. 
\item Randomly select the incidence of the model.   One variable is selected for a potential update, in line with a typical Metropolis scheme.
\item Compute the energy difference $\Delta E$ resulting from flipping the selected incidence variable.   This step quantifies how the configuration's energy would change if the variable were flipped.
\item Update the variable with probability $P = e^{-\beta \Delta E}$ if $\Delta E \geq 0$, and with $P = 1$ if $\Delta E < 0$. Energy-lowering moves are always accepted; energy-increasing ones are accepted with a probability based on $\Delta E$.
\item Repeat steps 2–4 for $n_{\text{steps}}$ iterations. The process continues for a predefined number of steps to ensure adequate mixing of the state space.
\end{enumerate}

\begin{figure}[t]
\centering
\includegraphics[width=\columnwidth]{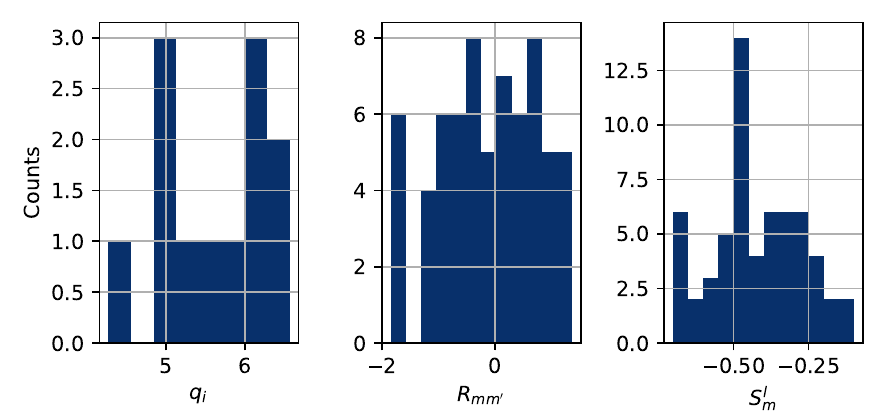}
\caption{Frequency of QUBO parameter strengths of the learned, statistical model}
\label{fig:model:qubo_hist}
\end{figure}

\begin{figure}[t]
\centering
\includegraphics[width=1.2\columnwidth]{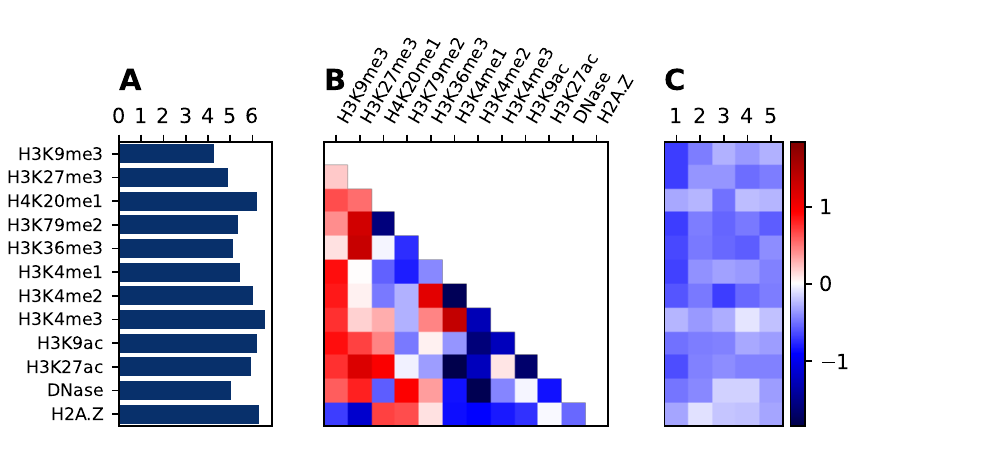}
\caption{ QUBO parameters visualized in separate categories. (A)  Each column represents an epigenetic marker, and the bar height shows its mean incidence across nucleosome positions.(B) 
Each column corresponds to an epigenetic marker pair, and the bar height indicates the intra-nucleosome coupling strength between those two markers at the same position.(C) 
Each column represents an epigenetic marker, and the bar height indicates the inter-nucleosome correlation strength for that marker across chromatin positions. These values quantify the extent to which the same marker tends to co-occur across different nucleosome sites.}
\label{fig:model:parameters}
\end{figure}

The learning progression for a typical optimization process of the iterative classical Boltzmann sampling is shown in Figure~ \ref{fig:model:errors}. The optimized parameters resulting from this learning process are shown in Figure~\ref{fig:model:parameters}. The frequency distribution of the QUBO parameter strengths generated using this approach is shown in Figure~\ref{fig:model:qubo_hist}.


\subsection{Mapping on A Quantum Annealer}
\label{sec:dwave_embedding}

The D-Wave quantum processing unit (QPU) is designed to address combinatorial optimization problems by physically implementing the quantum annealing paradigm \cite{King:2022aa,king2024computational}. At its core, the QPU comprises a lattice of superconducting qubits with low error rate \cite{Ansari_2016,Ansari_2013,Xu_2024,McBroom_Carroll_2024,Xu_2026}, each constructed from Josephson-junction-based circuits. These circuits operate at millikelvin temperatures within a dilution refrigerator, a cryogenic environment that suppresses thermal noise and helps maintain quantum coherence \cite{yarkoni2022quantum}. To further reduce decoherence, the qubits are shielded against external electromagnetic interference, background radiation, and nonequilibrium quasiparticle poisoning \cite{ansari2013effect,ku2020suppression}.

Quantum annealing on this platform encodes the optimization problem into the ground state of an Ising Hamiltonian or, equivalently, into a QUBO model. The physical layout of the qubits follows a predefined hardware topology, which specifies the permitted pairwise couplings and consequently restricts the class of problem graphs that can be directly embedded. The QUBO model of Eq. (\ref{eq.Hx}) can be mapped into the following quantum Ising model:  
\begin{equation}
\hat{\mathcal{H}}_O = \sum_i h_i \hat{Z}_i + \sum_{i>j} J_{ij} \hat{Z}_i \hat{Z}_j
\label{eq:qa:ising_objective}
\end{equation}
The $\hat{Z}$ operator being the Pauli $Z$ matrix $\sigma^z$ and $h_i$ and $J_{ij}$ are the  external magnetic field and qubit-qubit coupling parameters, respectively.  

The QUBO framework of Eq. (\ref{eq.Hx}), or in a simpler format the objective QUBO function $H_O(x) = \sum_{i \geq j} Q_{ij} x_i x_j$ is equivalent to the Ising model by the transformation $x_i = (1 \pm Z_i)/2$ with  $Z_i$ being the state of $i$-th qubit, more precisely the eigenvalue of the Pauli $\hat{Z}$ operator on the qubit state. Therefore, the mapping on a quantum annealer requires the following interpretations, summarized in Table \ref{table:qa:qubo_ising_hamiltonian}.

\begin{table*}[ht]
\centering
\begin{tabular}{ |c||c|c|c||c| } 
\hline
Classical Feature & QUBO & Ising & Hamiltonian & Quantum Feature \\ 
\hline
\hline
Inactive Value & 0 & -1 & -1 & 1st $\sigma^\text{z}$ EV \\
\hline
Active Value & 1 & +1 & +1 & 2nd $\sigma^\text{z}$ EV \\
\hline
Variable & Bit & Spin & Qubit & Observable \\
\hline
Symbol & $x_i$ & $\sigma_i$ & $\hat{\sigma}_i$ & Symbol \\
\hline
Parameters & $Q$ & $h$, $J$ & $h$, $J$ & Parameters \\
\hline
\end{tabular}
\caption{Comparison of corresponding features in the QUBO and Ising models as well as the mapping to an objective Hamiltonian.}
\label{table:qa:qubo_ising_hamiltonian}
\end{table*}

During an annealing run, the system evolves from an initial Hamiltonian; typically dominated by a transverse field $\sum_{i} \hat{\sigma}^x_i$ with a simple, well-defined ground state toward a final Hamiltonian encoding the problem-specific cost function. The transverse field creates a high initial tunneling probability, so that the system can get out of the local energy minima of computational states. If the evolution proceeds sufficiently slowly, the adiabatic theorem suggests that the system will remain close to its instantaneous ground state throughout the process \cite{crosson2021prospects, AdiabaticUnlikelySpeedup}, enabling it to approximate the optimal solution at the end of the schedule.

\subsubsection{Variable Mapping}

In the context of modeling epigenetic chromatin domains, the underlying optimization problem—such as inferring domain boundaries or interactions based on marker co-localization is formulated as a QUBO problem, as described above. To solve this model using quantum annealing, it must be embedded onto the physical topology of the quantum annealer. Graph topologies impose constraints on which qubits (representing binary variables) can be directly coupled. Embedding involves a minor-embedding process where logical variables are mapped to chains of physical qubits connected by strong ferromagnetic couplings, preserving the intended logical connectivity. For chromatin models with complex spatial or interaction graphs—e.g., based on nucleosome contact maps or epigenetic co-modification networks—efficient embedding depends critically on graph sparsity, modularity, and compatibility with the native hardware connectivity. Reducing chain lengths and auxiliary qubit usage is crucial to mitigate noise and improve fidelity in mapping these biological structures onto quantum hardware.

We label the incidence matrix (qubits on the lattice in the Ising alternative picture) by a 1D string.   This is achieved by mapping the marker-index $m$ and nucleosome index $n$ onto the qubit/node index $i$ as
\begin{equation}
i(m,n) = n \times M + m
\label{eq:model:index_mapping}
\end{equation}
and since $m<M$ applies for all valid $m$, the inverse mapping follows trivially as
\begin{equation}
m(i) = i \, \mathbf{mod} \, M \quad \text{and} \quad n(i) = i \, \mathbf{div} \, M \text{.}
\label{eq:mode:inverse_index_mapping}
\end{equation}
Within modular ({\bf mod}) mathematics, {\bf div} refers to integer division, which gives the quotient (whole number part) of a division, ignoring the remainder.

This mapping ensures that the inner iterations concerns the markers and the outer iterations concerns the nucleosomes, so that variables of the same nucleosome but different markers will directly follow each other.

We can now label these qubits. The classical QUBO variables $x_m^n$, with $m$ being for marker, $n$ for nucleosome location,  are mapped to the following 1D array of qubit states $q_i$'s: $
\left[Z_0^0, Z_1^0, Z_2^0, \dots, Z_{M-1}^0, Z_0^1, \dots, Z_0^{N-1}, \dots, Z_{M-1}^{N-1}\right]$.

\begin{figure}[h!]
\centering
\includegraphics[width=\columnwidth]{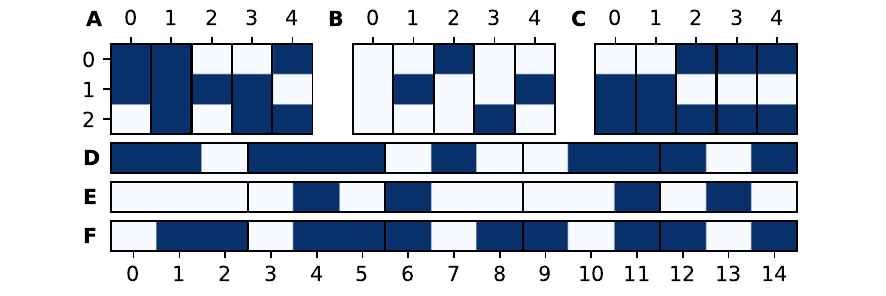}
\caption{Mapping between the incidence matrix of epigenetic markers and the qubits of the quantum annealing sampling  device, illustrated through three randomly initialized state examples of size $[3,5]$. Panels (A)-(C) show incidence matrices with three markers per nucleosome (Y axis) across five nucleosome positions (Y axis), while panels (D)-(F) show the corresponding linear qubit mappings on the quantum hardware (X labels correspond to individual qubits). E.g. the incidence at position "2,0" in panel (A) corresponds to qubit "6" in panel (D).Panels~(D)-(F) depict the linear indexing of the logical QUBO variables prior to minor embedding; they are not representations of the hardware graph or of the physical qubits used after embedding.}
\label{fig:model:incidence_qubit_mapping}
\end{figure}

Figure~\ref{fig:model:incidence_qubit_mapping}A–C illustrates three examples of incidence matrices, each representing three markers for five nucleosomes separated horizontally. A filled box denotes an incidence value of 1, while an empty box represents 0. Panels D–F in the same figure show the corresponding mappings of these incidence variables onto the physical qubit index, representing the three examples in the vectorized format suitable for being addressed by qubit indices. 
 It is important to note that Figure \ref{fig:model:incidence_qubit_mapping} illustrates the \emph{logical} mapping used to index the QUBO variables before embedding. Panels~D--F therefore show a 1:1 correspondence between logical variables and their linear indices, \emph{not} a representation of the physical qubits used after minor embedding. The actual hardware embedding typically maps each logical variable to a chain of physical qubits. The linear layout in \ref{fig:model:incidence_qubit_mapping} is used to visualize the flattening of the incidence matrix and is not intended to depict the connectivity structure of the problem or of the hardware graph.

\subsubsection{Hardware Embedding}

\begin{figure}[h!]
\centering
\includegraphics[width=1.1\columnwidth]{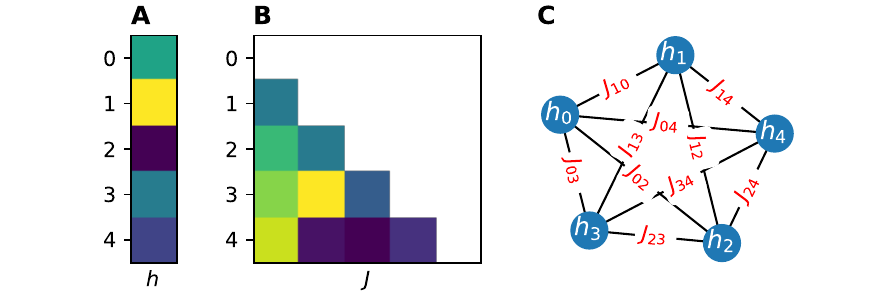}
\caption{Mapping of Ising parameters onto an objective graph. (A) Linear tuning of individual Ising spins as vector $h$. (B) Quadratic coupling between two Ising spins as a matrix $J$. (C) Representation of all Ising parameters as an objective graph where $J$ is the weighted adjacency matrix of that graph. Colors in (A) and (B) denote different parameter values.}
\label{fig:qa:example_ising_graph}
\end{figure}

The objective Hamiltonian consists of local field terms acting on individual qubits and pairwise ZZ interactions between qubit pairs. This Hamiltonian can be naturally represented as a weighted, undirected graph—see Fig.\ref{fig:qa:example_ising_graph} for an illustrative example. In this representation, each qubit (or binary variable) corresponds to a node in the graph. The local energy biases (or Z-field terms), denoted by h, are assigned as weights to the nodes, while the two-qubit interaction strengths J (i.e., ZZ couplings \cite{xu2021zz,xu2024lattice}) are mapped to weighted edges between the respective nodes. Specifically, the h values shown in Fig.\ref{fig:qa:example_ising_graph}(A) and the J couplings in Fig.\ref{fig:qa:example_ising_graph}(B) are jointly encoded in the graph structure presented in Fig.\ref{fig:qa:example_ising_graph}(C). Notably, the matrix J corresponds to the weighted adjacency matrix of the interaction graph. Because each node represents a binary variable, the measured eigenvalues of the Pauli-Z operator for each qubit directly determine the value of the corresponding variable in the final readout.

{\bf Required Connectivity in the Graph:} In general, all markers andnucleosomee locations are allowed to be coupled in a general matrix, which means that the objective graph is potentially fully connected (a complete graph), but the Cartesian nature of modelling indicates that lower (or sparse) connectivity is more practical, but not too low! How much of connectivity is sufficient for embedding a problem on a graph?   A quantitative measure of connectivity is sensible for the analysis of different task objectives. 

Consider a graph $G$ with $\mathbf{E}(G)$ being the set of all edges and $\mathbf{V}(G)$  being the set of all nodes. The number of all edges is $|\mathbf{E}(G)|$ and the number of nodes $\left| \mathbf{V}(G) \right|$. For this graph, a quantitative measure of connectivity is the average degree $\langle k \rangle_G$ of nodes, defined as the average number of edges that each node possesses, given by $\langle k\rangle_G = 2 \cdot \left| \mathbf{E}(G) \right| / \left| \mathbf{V}(G) \right|$. 

Another measure is gamma $\gamma_G$ which is defined as $\gamma_G\equiv 2 \cdot |\mathbf{E}(G)|/(|\mathbf{V}(G) \cdot (\mathbf{V}(G))$. The denominator is, in fact, the maximum number of edges in a complete graph with the same number of nodes. The interpretation of $\gamma_G$ is the following: for a fully connected (complete) graph $\gamma_G=1$, for a very sparse graph $\gamma_G\to 0$. Higher $\gamma_G$ means a more complex embedding is required on sparse hardware.

{\bf Chromatin Domain as an Objective Graph:}  We now analyze the structure of the objective graph associated with the chromatin domain model. In this context, each binary variable represents the presence or absence of a specific epigenetic marker at a given nucleosome location. Without internucleosomic coupling, the resulting graph is a \emph{cluster graph}, that is, a disjoint union of complete subgraphs~\cite{ClusterGraphs}. Each cluster corresponds to a nucleosome, where all epigenetic markers at that location are fully connected due to their internal interactions.

\begin{figure}[h!]
\centering
\includegraphics[width=\columnwidth]{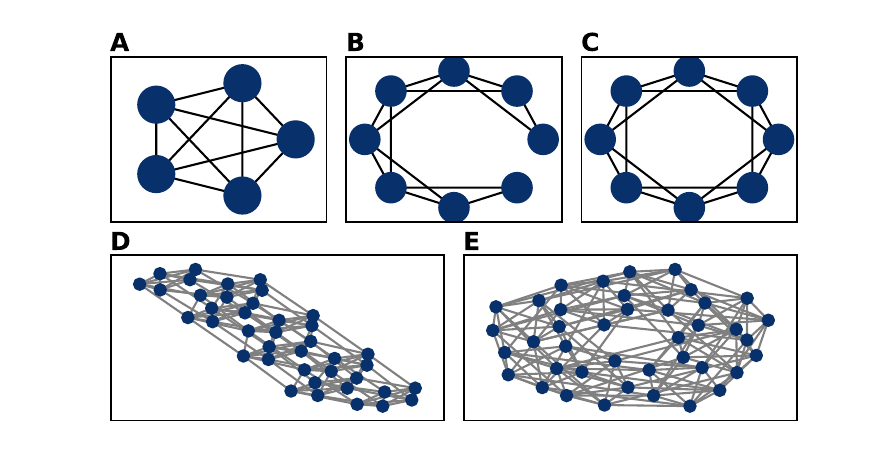}
    \caption{Structure of the objective graph under different boundary conditions. (A) A  marker intersection graph \( \mathbf{I}_M \): a complete graph encoding intra-nucleosomic marker incidences. (B) A nucleosome intersection graph \( \mathbf{I}_N \) with non-periodic boundaries: a \( 2L \)-regular graph representing inter-nucleosomic correlations. (C) Same as (B), but with periodic boundaries. (D) Full objective graph constructed as the Cartesian product of (A) and (B). (E) Same as (D), but using periodic boundaries from (C). Linear biases and coupling strengths are omitted for clarity.}
\label{fig:model:embed:cartesian_intersection_complete}
\end{figure}

To gain a deeper understanding of the structure of this model, we consider two \emph{intersection graphs}, constructed from sets of edges that encode marker-specific or nucleosome-specific interactions\footnote{In general, an intersection graph is a graph where each node represents a set, and an edge exists between two nodes if the corresponding sets have a non-empty intersection~\cite{IntersectionGraphs}. In our setting, sets are defined by the connectivity of markers or nucleosomes.}:

\begin{enumerate}
    \item \textbf{Marker intersection graph} \( \mathbf{I}_M \): This graph is constructed by grouping all edges associated with the same nucleosom but between all different markers. Since each marker is allowed to interact with every other marker within the same nucleosome, the resulting graph is a complete graph \( K_M \), isomorphic to the intra-nucleosomic subgraphs of the full objective graph (see Fig.~\ref{fig:model:embed:cartesian_intersection_complete}A).

    \item \textbf{Nucleosome intersection graph} \( \mathbf{I}_N \): This graph is built from sets of edges associated with each nucleosome's connections to others. For a model with local inter-nucleosomic interactions (up to a fixed distance \( L \)), the graph becomes a cyclic \( 2L \)-regular graph (see Fig.~\ref{fig:model:embed:cartesian_intersection_complete}B). In the full-scale configuration, this structure leads to a connectivity index of \( \gamma_G \approx 0.07 \), indicating a sparse coupling topology.
\end{enumerate}

The complete objective graph is given by the Cartesian product \( G_O = \mathbf{I}_M \mathsmaller{\square} \mathbf{I}_N \), as shown in Fig.~\ref{fig:model:embed:cartesian_intersection_complete} D- E. 

Since the models under consideration are restricted to Cartesian coupling structures, i.e., there are no mixed interactions between marker and nucleosome indices, this graph construction holds uniformly across all variants.

When embedding the problem graph on D-Wave hardware, it is often necessary to represent a single logical variable using a \textbf{chain} of multiple physical qubits. These qubits are strongly coupled to ensure that they behave as one unit and return the same value (ideally) after annealing. This chaining becomes essential when the hardware graph lacks direct connections between certain logical neighbors in the problem. To ``bridge'' such gaps, longer chains of connected physical qubits must be created.

\emph{Long chains are problematic}: they are more likely to break during annealing, meaning that not all qubits in the chain agree on a single value. Additionally, longer chains consume more physical qubits, thereby reducing the number of logical variables that can be embedded within the hardware's capacity. For example, embedding a fully connected graph of 20 variables (that is, a clique \( K_{20} \)) on the sparse hardware of D-Wave requires simulating many logical connections using chains. This can result in hundreds of physical qubits being tied up solely to represent a small set of logical variables.  In section (\ref{sec locality}) we examine the characteristics of these qubit chains in more detail.

Although the objective graphs in most practical optimization tasks are not fully connected, their required connectivity often exceeds the native connectivity available on current quantum annealing chips. This discrepancy becomes even more pronounced as the size of the objective graph \( G_O \) grows. As a result, direct implementation of such problem graphs onto existing hardware is generally infeasible without substantial embedding overhead.

Further analysis is presented in Supplementary Material \cite{kempe2025}, where we examine the embeddability of a specific instance with 12 epigenetic markers, 25 nucleosome positions, and a maximum interaction range of 5, under periodic boundary conditions. The model is mapped onto various hardware topologies, and we evaluate key embedding metrics such as chain length and total qubit overhead.  After describing how we process the epigenomic data in Supplementary Material \cite{kempe2025} we discuss the problem of embedding the objective graph in the Pegasus and Zephyr topologies in Supplementary Material \cite{kempe2025}  and the boundary condition and coupling threshold in Supplementary Material \cite{kempe2025}. 

\subsubsection{Embedding Locality}
\label{sec locality}

Due to the limited connectivity between qubits in quantum annealing hardware, it is important to understand how the structure of the problem graph, specifically, the couplings between variables, affects both the embedding process and the quality of the annealing outcome. A key factor in this analysis is the {\bf locality} of the required couplings, whether they are limited to nearest neighbors or to next to nearest neighbors, etc.  This is typically measured by the \textit{mean} and \textit{maximum chain lengths} resulting from the embedding. In this section, we explore these characteristics, focusing in particular on how they change with the maximum interaction range \( L \).

\begin{figure}[h!]
\centering
\includegraphics[width=\columnwidth]{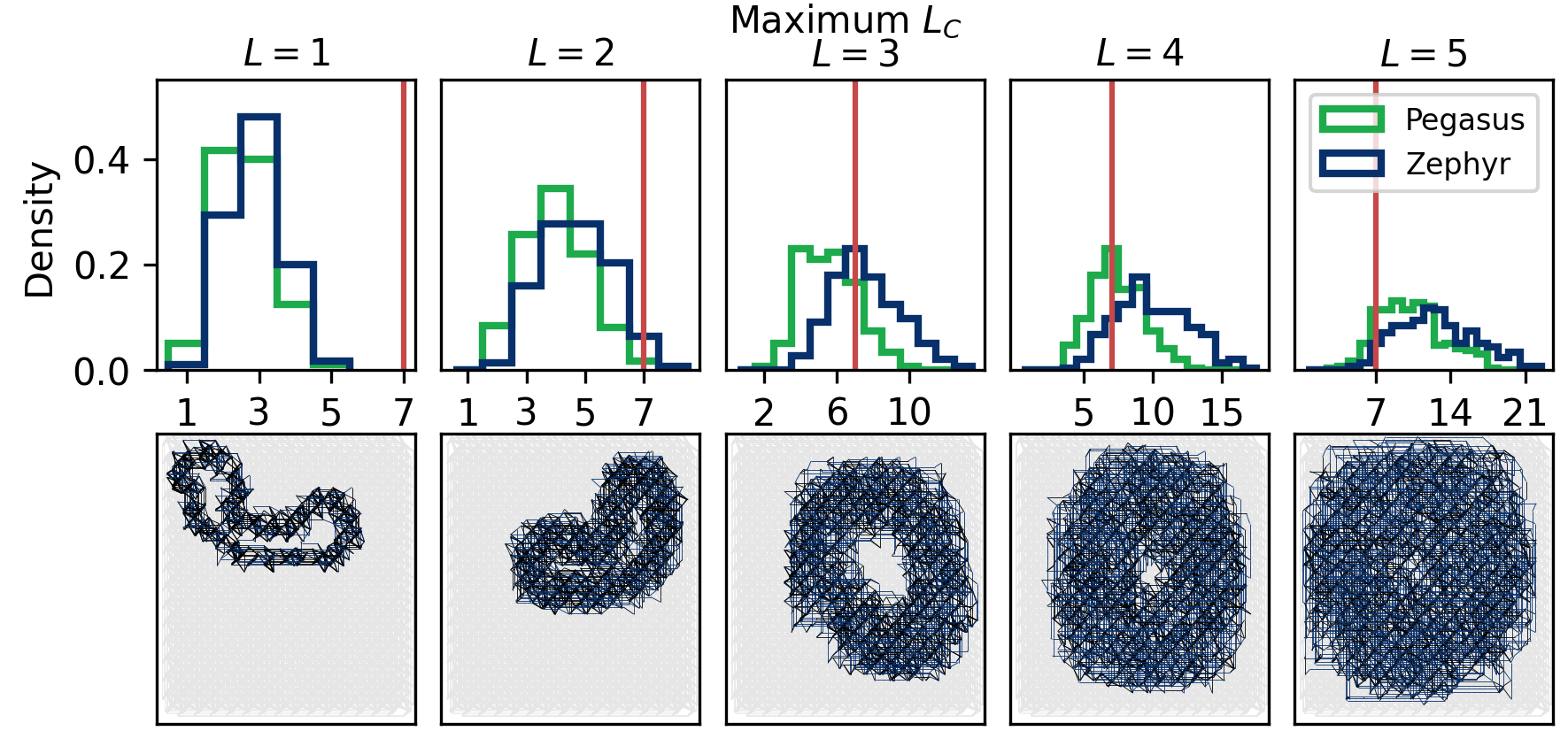}
\caption{Distribution of maximum chain lengths for model configurations $[12, 25, [1 \dots 5]]$. The recommended maximum chain length, $L_C = 7$, is shown as a red line. Below each distribution, the corresponding embedding on a Pegasus graph is shown.}
\label{fig:model:embed:long_range_chain_length}
\end{figure}

We embed some Chromatin states with a fixed number of epigenetic markers \( M \), nucleosome positions \( N \) and inter-nucleosome interaction range \( L \) on Pegasus and Zephyr graphs. In each case, depending on $L$, we notice a maximum chain length of a certain embedding.  We can determine the embedding for the chromatim model of $M=12$, $N=25$, and $L=\{1,2,3,4,5\}$. Figure.~\ref{fig:model:embed:long_range_chain_length} shows the distribution of {\bf chain lengths} required to embed these models on the hardware for different $L$. In these plots x-axis is the chain length in the hardware graph and y-axis indicates the normalized distribution of each chain length. We ignore the single-qubit (zero chain) from the distribution. We also consider that chain length 1 means one physical qubit to represent one incidence in the QUBO lattice.  Moreover, it was recommended to consider the maximum chain length of $L_C = 7$, which we show in red lines. Under each distribution, one can see the corresponding embedding on a Pegasus graph. 

By growing the range of internucleosome interaction $L$ in the chromatin states, the chain distribution in the hardware graph moves towards a larger peak. This growing complexity is also evident in the embedded graph visualizations shown in the lower panels of Figure.~\ref{fig:model:embed:long_range_chain_length}. Although all examples encode the same number of logical qubits, the number of physical qubits required varies significantly. This difference arises from the increasing spatial reach of the interactions in the objective graph as \( L \) grows, which demands longer chains and higher qubit overhead during embedding. Our analysis shows that only for the nearest-neighbor interaction range ($L=1$) in the nucleosomes does the entire chain distribution fall below the hardware upper limit $L_C$. Also, keeping the interaction range at $1< L\leq 3$ makes it possible to have the majority (but not all) of chains shorter than $L_C$. This is in agreement with the results in the Supplementary Material \cite{kempe2025}, in which the maximum correlation length plays a key role in embedding complexity for large-scale models.

In addition to the standard definition of chain length (the number of physical qubits comprising a chain), a second metric, the \emph{chain diameter} \( D_C \), is also a useful quantity.  It is defined by considering the number of qubits in the longest path of a chain sub-graph.\footnote{Note that the typical graph theory definition of a graph diameter refers to the longest path, not the number of nodes in the longest path (i.e. $+1$).} 

\begin{figure}[h!]
\centering
\includegraphics[width=\columnwidth]{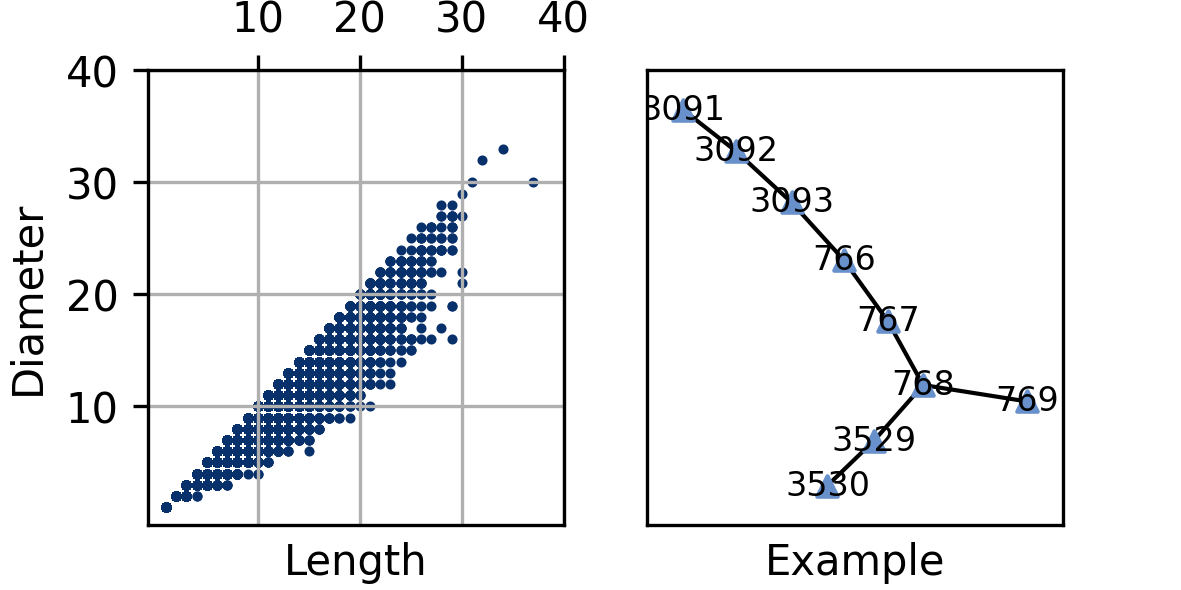}
\caption{Relation between chain lengths and chain diameters (left). Example of a qubit chain with different lengths and diameters, with denoted qubit indices on the Pegasus topology (right).}
\label{fig:model:embed:chain_diameter}
\end{figure}

An illustrative example of a chain with $D_C \neq L_C$ is given in figure \ref{fig:model:embed:chain_diameter} along with the distribution of $D_C$ and $L_C$ for numerous embeddings of models of different sizes. In the most extreme case, the chain diameter is $60\,\%$ smaller than the associated chain length. This illustrates that chains are not solely used to bridge distances in the target graph, but also to create effective logical sub-graph structures that otherwise do not exist in the target graph.

\section{Conclusion}

This study demonstrates the potential of quantum annealing for sampling biologically realistic chromatin states using learned model parameters. Our approach does not compute explicit architectural metrics such as TAD size distributions or insulation scores. Instead, it focuses on reproducing statistical features such as mean marker incidences and intra-inter-nucleosome correlations. We generate configurations that exhibit TAD-like structural motifs. The chromatin domain sampling model is particularly well-suited for Quantum Annealing Sampling (QAS), due to its low graph connectivity and orthogonal layout of the objective graph.

We observe favorable empirical embedding scalability when embedding the model on current D-Wave hardware architectures, in the sense that required chain lengths and qubit overhead grow moderately across the biologically relevant range of model sizes considered. Here, scaling refers to empirical embeddability metrics derived from the embedding statistics (e.g., chain length and qubit overhead), rather than to asymptotic or algorithmic scaling behavior. These trends are supported by the embedding analyses shown in Figure~\ref{fig:model:embed:long_range_chain_length} and Figure~\ref{fig:model:embed:grid_sweep}.
Although the limited number of qubits imposes constraints for larger datasets, the introduction of a coupling-strength threshold offers a practical solution: it reduces embedding complexity while preserving model performance. Notably, the Zephyr topology used in the D-Wave Advantage2 system allows for lower-complexity embeddings compared to the older Pegasus topology, in agreement with prior observations in combinatorial optimization tasks.

Our analysis of annealing parameters shows that sampling tasks require different optimal settings, where longer annealing times and stronger chain strengths often improve performance, sampling benefits from intermediate values that preserve diversity in the sampled states. This distinction highlights the need for tailored quantum schedules when applying annealing to biological sampling tasks.
We further validate our approach by comparing quantum and classical sampling statistics across multiple model sizes. Future work will extend this framework to derive interpretable architectural metrics, enabling direct comparison with experimental TAD boundaries and insulation profiles.
These results provide a proof-of-conceproof of conceptf quantum hardware in chromatin sampling workflows. Additionally, we show that incorporating empirical bias, either through direct incidence encoding or reverse annealing, enables targeted sampling near biologically relevant states.

Importantly, quantum annealing sampling exhibits distinct qualitative timing characteristics arising from hardware-level parallelism. For sufficiently small models, multiple independent sampling Hamiltonians can be embedded and sampled concurrently on a single annealing chip, enabling the generation of many samples within a single anneal. Once a fixed embedding is computed, its one-time cost can be amortized across many annealing readouts. 
We emphasize that this observation reflects workflow‑level throughput and hardware-level parallelism, meaning that multiple independent sampling instances can be embedded and sampled concurrently within a single anneal, thereby increasing sample throughput without reducing the runtime of an individual anneal. This should not be interpreted as a quantitative runtime or asymptotic speedup.

Taken together, our findings suggest that quantum annealing provides a promising route for sampling chromatin configurations consistent with empirical epigenetic statistics and capable of exhibiting TAD-like structural features. This represents a step toward leveraging near-term quantum hardware for interpretable modeling of epigenetic regulation.

\section*{Acknowledgements}

We gratefully acknowledge the support of the JSC department at Forschungszentrum Jülich and D-Wave Systems for providing access to the quantum annealing resources required to carry out the experiments presented in this work. We thank Giorgi Kvantrishvili for comments and  feedback.

\section*{Funding Declaration} This research was partially funded by the Horizon Europe project OpenSuperQPlus100 under Grant Agreement No.~101113946, and National Science Foundation Grant No. 2242763.

\section*{Author Contribution}
T.K. carried out the numerical analysis and prepared the plots, under the formal supervision of M.A. and co-advising of A.T. M.A. with the assistance of A.T. wrote the main manuscript. A.T. contributed biophysics expertise and provided advisory input. All authors reviewed and approved the final manuscript.

\section*{Data availability}
 The datasets analysed during the current study are available to download from the Roadmap Epigenomics Project at \url{https://egg2.wustl.edu/roadmap/web_portal/} (see \cite{RoadmapEpigenomicsProject}) in \texttt{.bigwig} format. It was compiled for further analysis by the \texttt{bigWigToBedGraph} executable, obtained from the binary utilities directory of UCSC: \url{http://hgdownload.soe.ucsc.edu/admin/exe/linux.x86_64/} (see \cite{BinaryUtilitiesDirectory}). We collected our codes relevant to the results of this paper in the following GitHub folder: \url{https://github.com/tobiaskempe/qas-cdf}.

\bibliographystyle{ieeetr}  
\bibliography{MAbib,references}

\end{document}